\documentclass[11pt,a4paper]{article}
\usepackage{epsfig}
\usepackage{amssymb}
\usepackage{graphicx}
\usepackage{color}
\usepackage{subfigure}

\usepackage{latexsym}

\begin{document}

\baselineskip 6mm
\renewcommand{\thefootnote}{\fnsymbol{footnote}}

\newcommand{\nc}{\newcommand}
\newcommand{\rnc}{\renewcommand}

%%%%%%%%%%%%%%%%%%%%%% Equation Numbering %%%%%%%%%%%%%%%%%%%%%%%
%\makeatletter \rnc{\theequation}{\thesection.\arabic{equation}}
%\@addtoreset{equation}{section} \makeatother

%%%%%%%%%%%%%%%%%%%%%%%%%%%%%%%%%%%%%%%%%%%%%%%%%%%%%%%%%%%%%%%%%
%                                                               %
%                NEW COMMANDS AND MACROS                        %
%                                                               %
%%%%%%%%%%%%%%%%%%%%%%%%%%%%%%%%%%%%%%%%%%%%%%%%%%%%%%%%%%%%%%%%%

\newcommand{\tcb}{\textcolor{blue}}
\newcommand{\tcr}{\textcolor{red}}
\newcommand{\tcg}{\textcolor{green}}

%%%%% Simplify some frequently used LaTeX commands %%%%%

\def\beq{\begin{equation}}
\def\eeq{\end{equation}}
\def\ba{\begin{array}}
\def\ea{\end{array}}
\def\bea{\begin{eqnarray}}
\def\eea{\end{eqnarray}}
\def\nn{\nonumber}

%%%%%%%%%%%%%%%%%%%%%%%%%%%%%%%%%%%%%
%  Journal 
%%%%%%%%%%%%%%%%%%%%%%%%%%%%%%%%%%%%

\def\CMP{Commun. Math. Phys.~}
\def\JHEP{JHEP~}
\def\Pre{Preprint}
\def\PRL{Phys. Rev. Lett.~}
\def\PR {Phys. Rev.~}
\def\CQG {Class. Quant. Grav.~}
\def\PL {Phys. Lett.~}
\def\NP {Nucl. Phys.~}

%%%%%%%%%%%%%%%%%%%%%%%%%%%%%%%%%%%%%%%%%%%%%%%%%%
%   Boldface Letters
%%%%%%%%%%%%%%%%%%%%%%%%%%%%%%%%%%%%%%%%%%%%%%%%%%
\def\G{\Gamma}

\def\S{{\bf S}}
\def\C{{\bf C}}
\def\Z{{\bf Z}}
\def\R{{\bf R}}
\def\N{{\bf N}}
\def\M{{\bf M}}
\def\P{{\bf P}}
\def\bm{{\bf m}}
\def\bn{{\bf n}}

\def\CA{{\cal A}}
\def\CB{{\cal B}}
\def\CC{{\cal C}}
\def\CD{{\cal D}}
\def\CE{{\cal E}}
\def\CF{{\cal F}}
\def\CM{{\cal M}}
\def\CG{{\cal G}}
\def\CI{{\cal I}}
\def\CJ{{\cal J}}
\def\CL{{\cal L}}
\def\CK{{\cal K}}
\def\CN{{\cal N}}
\def\CO{{\cal O}}
\def\CP{{\cal P}}
\def\CQ{{\cal Q}}
\def\CR{{\cal R}}
\def\CS{{\cal S}}
\def\CT{{\cal T}}
\def\CV{{\cal V}}
\def\CW{{\cal W}}
\def\CX{{\cal X}}
\def\CY{{\cal Y}}
\def\We{{W_{\mbox{eff}}}}

%%%%%%%%%%%%%%%%%%%%%%%%%%%%%%%%%%%%%%%%%%%%%%%%%%
%   Mathematical Symbols  AA
%%%%%%%%%%%%%%%%%%%%%%%%%%%%%%%%%%%%%%%%%%%%%%%%%%

\newcommand{\p}{\partial}
\newcommand{\bp}{\bar{\partial}}

\newcommand{\half}{\frac{1}{2}}

\newcommand{\bfalpha}{{\mbox{\boldmath $\alpha$}}}
\newcommand{\bfbeta}{{\mbox{\boldmath $\beta$}}}
\newcommand{\bfgamma}{{\mbox{\boldmath $\gamma$}}}
\newcommand{\bfmu}{{\mbox{\boldmath $\mu$}}}
\newcommand{\bfpi}{{\mbox{\boldmath $\pi$}}}
\newcommand{\bfvarpi}{{\mbox{\boldmath $\varpi$}}}
\newcommand{\bftau}{{\mbox{\boldmath $\tau$}}}
\newcommand{\bfeta}{{\mbox{\boldmath $\eta$}}}
\newcommand{\bfxi}{{\mbox{\boldmath $\xi$}}}
\newcommand{\bfkappa}{{\mbox{\boldmath $\kappa$}}}
\newcommand{\bfepsilon}{{\mbox{\boldmath $\epsilon$}}}
\newcommand{\bfTheta}{{\mbox{\boldmath $\Theta$}}}

\newcommand{\bz}{{\bar{z}}}

\newcommand{\dalpha}{\dot{\alpha}}
\newcommand{\dbeta}{\dot{\beta}}
\newcommand{\blambda}{\bar{\lambda}}
\newcommand{\btheta}{{\bar{\theta}}}
\newcommand{\bsigma}{{{\bar{\sigma}}}}
\newcommand{\bepsilon}{{\bar{\epsilon}}}
\newcommand{\bpsi}{{\bar{\psi}}}

%%%%%  Temporary notation %%%%

\def\ct{\cite}
\def\la{\label}
\def\eq#1{(\ref{#1})}

%%% Greek letters %%%

\def\a{\alpha}
\def\b{\beta}
\def\g{\gamma}
\def\G{\Gamma}
\def\d{\delta}
\def\D{\Delta}
\def\ep{\epsilon}
\def\e{\eta}
\def\ph{\phi}
\def\Ph{\Phi}
\def\ps{\psi}
\def\Ps{\Psi}
\def\k{\kappa}
\def\l{\lambda}
\def\L{\Lambda}
\def\m{\mu}
\def\n{\nu}
\def\th{\theta}
\def\Th{\Theta}
\def\r{\rho}
\def\s{\sigma}
\def\S{\Sigma}
\def\ta{\tau}
\def\o{\omega}
\def\O{\Omega}
\def\pr{\prime}

%%%%% Mathematical Symbols

\def\half{\frac{1}{2}}

\def\goto{\rightarrow}

\def\na{\nabla}
\def\grad{\nabla}
\def\curl{\nabla\times}
\def\div{\nabla\cdot}
\def\pa{\partial}

\def\bra{\left\langle}
\def\ket{\right\rangle}
\def\lb{\left[}
\def\lc{\left\{}
\def\ls{\left(}
\def\lp{\left.}
\def\rp{\right.}
\def\rb{\right]}
\def\rc{\right\}}
\def\rs{\right)}
\def\cl{\mathcal{l}}

\def\vac#1{\mid #1 \rangle}

%%%%  Special symbol
\def\td#1{\tilde{#1}}
\def\check{ \maltese {\bf Check!}}

%%%%% Roman pont in math

\def\Tr{{\rm Tr}\,}
\def\det{{\rm det}\,}

%%%%% Special format

\def\bc#1{\nnindent {\bf $\bullet$ #1} \\ }
\def\ch {$<Check!>$ }
\def\ss {\vspace{1.5cm}}

\begin{titlepage}

%---------------- preprint number ---------------
\hfill\parbox{5cm} { }

\hskip1cm

 \hskip12cm{CQUeST--2013-0590}

\vspace{10mm}

\begin{center}
%------------------------ title ------------------------
{\Large \bf Conserved quantities and Virasoro algebra in New massive gravity }

%---------------- authors and addresses ----------------
\vskip 1. cm
  {
  Wontae Kim$^{ab}$\footnote{e-mail : wtkim@sogang.ac.kr},  
  Shailesh Kulkarni$^a$\footnote{e-mail : skulkarnig@gmail.com} and 
  Sang-Heon Yi$^{a}$\footnote{e-mail : shyi@sogang.ac.kr} 
  }

\vskip 0.5cm

{\it $^a\,$Center for Quantum Spacetime (CQUeST), Sogang University, Sinsu-dong, Mapo-gu, Seoul 121-742, Korea}\\
{\it $^b\,$Department of Physics,, Sogang University, Sinsu-dong, Mapo-gu, Seoul 121-742, Korea}\\

\end{center}

\thispagestyle{empty}

\vskip1.5cm

%----------------------- abstract ----------------------

\centerline{\bf ABSTRACT} \vskip 4mm

\vspace{1cm}
\noindent Using the {\it off-shell} Noether current and potential we compute the entropy for the AdS black holes
in new massive gravity.  For the non-extremal BTZ black holes  by implementing the so-called stretched horizon approach
we reproduce the correct expression for the horizon entropy. For the extremal case, 
 we adopt standard formalism in the AdS/CFT correspondence and reproduce the corresponding  entropy by 
computing the central extension term on the asymptotic boundary of the near horizon geometry. We explicitly show the invariance of the angular momentum along the radial direction for extremal as well as non-extremal
BTZ black holes in our model.  Furthermore, we extend this invariance for the black holes in new massive gravity
coupled with a scalar field, which correspond to  the holographic renormalization group  flow trajectory
of the dual field theory. This provides another realization for the holographic c-theorem.  

\vspace{2cm}
%PACS numbers:

%\today

\end{titlepage}

\renewcommand{\thefootnote}{\arabic{footnote}}
\setcounter{footnote}{0}

%\tableofcontents
%%%%%%%%%%%%%%%%%%%%%%%%%%%%%%
%                                                                            %
%   Sec.  Introduction                                                       %
%                                                                            %
%%%%%%%%%%%%%%%%%%%%%%%%%%%%%%

%%%%%%%%%%%%%%%%%%%%%%%%%%%%%%%%%%%%%%%%%%%%%%%%%%%%%%
\section{Introduction}
%%%%%%%%%%%%%%%%%%%%%%%%%%%%%%%%%%%%%%%%%%%%%%%%%%%%%
Local symmetries play a crucial role in understanding the thermodynamical properties of black holes. 
It was shown that the Noether charges corresponding to the diffeomorphism symmetry of any generally covariant
theory of gravity are related to the black hole entropy \cite{Wald:1993nt,Iyer:1994ys}. This approach, when
applied to the conventional Einstein-Hilbert action, reproduce the well known  Bekenstein-Hawking 
entropy (BH) \cite{Bekenstein:1973ur, Hawking:1974rv}. However, the Wald's formalism tells us little about
the microscopic degrees of freedom responsible for the black hole entropy. A major step to understand
the black hole entropy from the microscopic point of view was taken    
%Symmetry has been one of the fundamental principles in modern physics. All the known 
%elementary interactions at the microscopic level are governed by some (local) symmetries: 
%gauge symmetry and diffeomorphism.  In the duality picture of  the more sophisticated 
%unifying  string/M-theory, it has been the first criterion for the duality maps among seemingly 
%remote    theories.  Whenever one is equipped with Lagrangian,  symmetry is usually associated 
%with the corresponding conserved current, as was discovered by Noether.  Symmetry principles 
%play  important roles in the black hole thermodynamics. Indeed, it was shown that
%the Noether charges corresponding to the diffeomorphism symmetry of any generally covariant
%theory of gravity are related to the black hole entropy \cite{Wald:1993nt,Iyer:1994ys}. Wald's approach when
%applied to the conventional Einstein-Hilbert Lagrangian, reproduce the well known  Bekenstein-Hawking 
%entropy (BH) \cite{Bekenstein:1973ur, Hawking:1974rv}. Perhaps the most interesting application of the symmetries is to uncover the microscopic 
%origin of the black hole entropy. The vert first attempt in this direction was made 
by Strominger and Vafa \cite{Strominger:1996sh}. They showed that certain extremal black holes in string theory can be described, 
through a string duality map, by two dimensional  conformal field theory (CFT).  Then the entropy of this 
CFT was obtained by using the Cardy formula \cite{Cardy:1986ie} and shown to be consistent with the 
standard BH entropy. In the related subsequent development \cite{Strominger:1997eq}, 
the entropy of the Banados-Teitelboim-Zanelli (BTZ) 
black holes~\cite{Henneaux:2002wm}  was also obtained through two dimensional CFT. This formalism relies on the 
remarkable observation that the algebra among the asymptotic symmetry generators
is isomorphic to two copies of the Virasoro algebra  with  the central charge $c$ \cite{Brown:1986nw}
 \beq
[Q_{m},Q_{n}]  =  (m-n)Q_{m+n} + \frac{c}{12}m(m^{2}-1)\delta_{m+n,0}
 \,. \label{eq:Brownalgebra}
%\lb\bar{L}_{m}, \bar{L}_{n}\rb &=& (m-n) \bar{L}_{m+n} + \frac{\bar{c}}{12}m^{3}\delta_{m+n,0}  
\eeq
This appearance of the Virasoro algebra of symmetry generators can be regarded as  the predecessor  of the AdS/CFT correspondence 
\cite{Maldacena:1997re}. 

An alternative approach which uses only the near horizon properties of black holes was given by Carlip 
\cite{Carlip:1998wz, Carlip:1999cy}. In this `stretched horizon' approach  one begins by assuming the existence of 
an approximate null Killing vector. The  location of the Killing horizon is determined by the 
vanishing of the norm of that Killing vector. Under certain boundary conditions near the Killing horizon, 
it was shown that  the Fourier modes of diffeomorphism generators realized  by vector fields form 
subalgebra isomorphic to Diff$(S^{1})$, which is also known as Witt algebra, as 
\beq
[ \xi_{m},\xi_{n} ]^{a} = -i (m-n)\xi^{a}_{m+n}. \label{eq:Diffalgebrafourier}
\eeq
The algebra among  canonical conserved 
charges corresponding to the above generators is identical with a copy of the standard Virasoro algebra.   
This construction of the algebra  indicates the existence of a certain 
two dimensional CFT, and allows us to read off the central charge of the underlying CFT. Then, 
the black hole entropy is reproduced by substituting the central charge and zero mode eigenvalue
of the conserved charge in the Cardy formula.  Based on this idea, several alternative methods  have been proposed 
\cite{Lin:1999gf,Brustein:2000fw, Solodukhin:1998tc, Jing:2000yd,
 Das:2000zs, Hotta:2000gx, Koga:2001vq, Park:2001zn, Silva:2002jq, Cvitan:2002cs, Kang:2004js, Cvitan:2004pd} 
to know more about the microscopic origin of the  horizon entropy.   

The above  stretched horizon approach has been revisited by Majhi {\it et.al.} \cite{Majhi:2011ws}
wherein the central charge and the horizon entropy is obtained by using the
 {\it off-shell} expressions for the Noether current and potential. One of the key features
of this method is that the on-shell vanishing part of the Noether current is not essential for  
performing calculations in brackets  among the Noether charges. In this sense the 
definition of bracket is more  general. This formalism extends in a 
straightforward manner to Lanczos-Lovelock models of gravity. 
Recently, it has been shown that the same expressions for the Noether charges
and horizon entropy can also be obtained by using either the Gibbons-Hawking
surface term or the gravitational surface term \cite{Majhi:2012tf}.
This analysis was based on the holographic property of the gravitational action functional \cite{Mukhopadhyay:2006vu}. 
Although  Lanczos-Lovelock models generalize  Einstein gravity to a great extent, 
 they still come under the two derivative theories.  
It is known that a specific combination of Ricci scalar and Ricci tensor leads 
to interesting gravity models in three and four dimensions~\cite{Bergshoeff:2009hq, Lu:2011zk}. 
The higher curvature gravity in $2+1$ dimensions, the so-called new massive gravity (NMG),  
is originally introduced as the parity even counter part  of topologically massive gravity~\cite{Deser:1981wh}. 
NMG  allows propagating massive gravitons and also incorporate various black hole solutions. Moreover, 
NMG has been shown to be consistent with the so-called holographic c-theorem~\cite{Freedman:1999gp, Sinha:2010ai,Myers:2010xs}.  
It is therefore natural to  extend the analysis of \cite{Majhi:2011ws} to obtain the
horizon entropy for black holes in genuinely higher derivative gravity
such as NMG. 

Another motivation to study $2+1$ dimensional Einstein gravity and 
NMG is as follows. For Einstein gravity, it was  shown
that the conserved quantities like mass and angular momentum for 
 Kerr black hole geometries in arbitrary dimensions  
take the same values on the near horizon and at the asymptotic 
infinity \cite{Julia:1998ys, Modak:2010fn}. 
A generalization of this result for the the asymptotically $AdS$ black holes by using the
improved surface integrals has been provided in \cite{Barnich:2003xg, Barnich:2004uw}. 
The improved surface integrals allow us to compute the conserved charges both at the horizon and 
the asymptotic infinity. In addition to this, it has also been shown that  
the angular momentum is  invariant not only 
just at two asymptotic boundaries but also all along the entire radial direction. 
This procedure is valid for  higher dimensional Kerr black holes in asymptotically flat 
as well as $AdS$ geometries.  

The rotating $AdS$ black holes in $2+1$ dimensions (BTZ black holes) 
possess an extra feature. Since the asymptotic as well
as the near horizon geometry correspond to  $AdS_{3}$ space, this geometry may be viewed as the holographic renormalization group (RG) flow between the
ultraviolate and the infrared CFT. 
This provides another realization of holographic c-theorem beyond domain wall solutions\footnote{The simplest 
examples of such holographic construction of RG flows are given by domain wall solutions 
interpolating two AdS spaces.}.
However, in the absence of any matter field this holographic RG flow becomes trivial.  The  next simplest 
step would be to study  the rotating black holes in $2+1$ dimensional Einstein gravity 
coupled to a scalar field. For the interpretation in the holographic RG flow,  the extremal black 
hole solutions are the relevant one \cite{Hotta:2008xt}\footnote{For some specific choice of 
the scalar potential the  non-extremal black hole solution is given in  \cite{Henneaux:2002wm, Martinez:2004nb, Correa:2012rc}.}. 
The deformed extremally rotating BTZ black holes in NMG coupled with a  scalar are also discussed in \cite{Kwon:2012zh}.
It was shown that the family of extremally rotating  hairy AdS black holes  can 
be described by reduced-order equations of motions(EOM). These black holes may be regarded 
as the scalar-hairy deformation of extremally rotating  BTZ black holes. 
However, the explicit derivation of the invariance of the angular momentum for these black holes has not been given
in the literature. The detailed analysis of the invariance of the angular momentum for the Einstein as well as NMG
by incorporating the scalar field would provide us fresh insights to understand  RG flow in the dual CFT. 

%enable us to understand the 
%
%
%
%
%  The BTZ black holes in this scalar-Einstein theory (deformed BTZ) 
%possess an extra feature. In this case the angular momentum is seen 
%to be invariant not only at just at two asymptotics boudaries but also all 
%along the entire radial direction. 
%Now, consider the extremally rotating deformed BTZ, wherein the near horizon 
%geometry is also $AdS_{3}$ . From the perspective of AdS/CFT correspondence
%the invariance of the conserved charge (angular momentum) along the radial 
%direction  would mean that  
%the scaling dimensions  of dual operators are invariant along the renormalization group
%(RG) flow between the ultraviolate (asymptotic $AdS_{3}$) and infrared (near horizon 
%$AdS_{3}$) conformal field thoeries. This provids another realization of holographic 
%c-theorem~\cite{Kwon:2012zh}. The NMG coupled to the scalar 
%field has been considered in \cite{Hotta:2008xt}.  
%However,  invariance of the angular momentum for this case has not been discussed in the 
%literature. The explicit derivation of invariace  the angular momentum 
%case would provide us a new direction to look at the holographic c-theorem for the
%higher derivative gravity such as NMG. 

The purpose of the present work is to compute the horizon entropy from the 
point of view of the  near horizon Virasoro algebra and to show explicitly the 
invariance of the angular momentum for the rotating BTZ black holes 
in NMG.  
%In this work, by using NMG as a toy example of higher curvature gravity, we study 
%the black hole entropy from the view point of near horizon Virasoro algebra.  
%To achive this we adopt the {\it off-shell} analysis of \cite{Majhi:2011ws}.   
We first consider the general expression for the {\it off-shell} Noether current and 
its potential \cite{Padmanabhan:2009vy, paddybook} for the NMG.  
%Then,  we  derive the expression for the entropy of  rotating BTZ black holes in NMG. 
 Then, by integrating the Noether potential on the stretched horizon we obtain the expression for the 
 Noether charge. Using the general definition of bracket given in \cite{Majhi:2011ws},  we construct
the algebra among the Noether charges. It turns out that this algebra is isomorphic to a copy of the Virasoro
algebra with a central extension. The zero mode eigenvalue and central charge are obtained by Fourier transforming 
the Noether charge and central extension term, respectively. 
Next, we evaluate these expressions for non extremal rotating BTZ black hole. 
Finally, using the  Cardy formula  we obtain the expression for the horizon  entropy. 
We also sketch a method to calculate the  entropy for extremal rotating BTZ black hole in NMG. 
 For this case we will proceed in a slightly different way.  We take the same expressions for the Noether current, 
potential and the bracket among the Noether charges mentioned earlier. However, we choose 
the diffeomorphisms  different from the non-extremal case, which preserves the fall-off boundary 
conditions \cite{Compere:2012jk} at the asymptotic infinity of the near horizon extremal BTZ. 
We show that algebra among the Noether charges is isomorphic to 
the Virasoro algebra with the central extension. The central charge identified from this algebra 
is consistent with the one given in \cite{Azeyanagi:2009wf}.

To show the angular momentum invariance for the rotating BTZ  black holes along the radial 
direction, we adopt a specific quasi-local method for conserved 
charges, which is known as  the Komar integrals. Here as well,  we use the same  Noether potential 
as in the computation for black hole entropy. We calculate the Noether potential corresponding to 
the rotational Killing vector for the BTZ black hole in the Einstein gravity and  NMG. 
The angular momentum is obtained by integrating this expression over the surface with 
codimension two situated at $ r_{H}\le r \le r_{\infty}$, where $r_{H}$ is the position of the outer horizon. 
In the Einstein case, our resulting expression for the angular momentum  matches exactly with the corresponding 
one given in \cite{Barnich:2003xg, Barnich:2004uw}. By applying the same technique  
we show the invariance of the angular momentum for the rotating BTZ black holes in NMG. 
As mentioned earlier, the invariance of the angular momentum for the rotating BTZ black holes  
is related to the holographic RG flow. However, for the pure Einstein or NMG theories this
RG flow is trivial. In order to gain further insights into the holographic RG flow, we consider extremally rotating 
BTZ solution for  NMG coupled with a scalar field. Then, we compute the relevant  Komar integrals \cite{Komar:1958wp}  
corresponding to the rotating Killing vector and show that the angular momentum is indeed invariant
along the radial direction.  

Apart from its relevance in understanding RG flow in the dual CFT's, the invariance of the angular momentum 
gives us an important relation between the infrared and ultraviolate entropies. 
According to the conventional AdS/CFT correspondence,  the central charge of the dual 
ultraviolet CFT is always greater than that of the infrared CFT, 
which is coined as holographic c-theorem.  Since the Cardy formula requires the conformal weights of dual states 
together with the central charge, this theorem is insufficient to identify which entropy of dual CFT corresponds to 
the  BH entropy of those black holes.  By establishing the angular momentum 
invariance of the hairy extremal black holes, we verify that  the entropy of the dual infrared CFT  is  
always less than or equal to the one of dual ultraviolet CFT and that the entropy of the infrared CFT  
gives the BH entropy of those black holes. This matching between the infrared dual CFT and the 
BH entropy of black holes is anticipated through the near horizon CFT approach~\cite{Carlip:1998wz}.   
However, we would like to emphasize that this near horizon CFT is not the same one with the infrared dual CFT 
used in holographic c-theorem.  Nevertheless, by using conserved currents related to the generalized Komar potential, 
one can see that  the same entropy of black holes can be reproduced in this way.

This paper is organized as follows. In the next section we summarize some basic facts about conserved currents  
for the local symmetry, for the sake of completeness, and then we consider  the 
modified Noether current which was introduced as the off-shell conserved currents~\cite{Padmanabhan:2009vy} 
and show that their meaning  may be understood as  the generalized Komar potential. 
 In section three we obtain the entropy of black holes through the near horizon dual CFT by using the 
{\it off-shell} formalism of \cite{Majhi:2011ws}. We conclude this section by providing 
a brief discussion on the entropy for extremally rotating BTZ black holes. 
In section four, using the generalized Komar potential, we obtain the quasi-local angular momentum of 
extremally rotating hairy AdS black holes on three dimensions and show its invariance along the radial direction.   
This verifies that the BH entropy of these black holes can be obtained by the infrared dual CFT.  
We summarize our results and discuss some future directions in the final section.  
Appendix-A contains some useful formulae and definitions in the stretched horizon approach.
A derivation of angular momentum invariance for the usual Einstein gravity is provided in Appendix-B.

%%%%%%%%%%%%%%%%%%%%%%%%%%%%%%%%%%%%%%%%%%%%%%%%%%
\section{Conserved currents and their potentials}\label{CCP}
%%%%%%%%%%%%%%%%%%%%%%%%%%%%%%%%%%%%%%%%%%%%%%%%%%
In this section we review and summarize the Noether procedure for   symmetry to fix our 
conventions. Since we are interested in the conserved charges and the entropy of black
 holes, it is useful to collect some well-known results for
 conserved currents for local symmetry to clarify our presentation (See for reviews, \cite{Compere:2007az, Szabados:2004vb, 
 Fischetti:2012rd}).

%%%%%%%%%%%%%%%%%%%%%%%%%%%%%%%%%%%%%%%%%%%%%%%%%%%%%%%%%%
%\subsection{Noether currents}\label{CCP1}
%%%%%%%%%%%%%%%%%%%%%%%%%%%%%%%%%%%%%%%%%%%%%%%%%%%%%%%%%
Let us consider the generic action, $I[\varphi]$, which enjoys local symmetry  and contains 
 various fields which are denoted collectively as $\varphi^i$. Some of these fields can be regarded as 
gauge fields and others as matter ones.
Under the generic variation of the field $\varphi\rightarrow \varphi +\delta \varphi$, 
the  action is varied as
\beq \delta I[\varphi] = \int d^d x \sqrt{-g}\Big[ E_i(\varphi)\delta \varphi^i + \nabla_{\mu}
\Theta^{\mu}(\varphi,\delta \varphi)\Big]\,,  \eeq
where $E_i(\varphi)=0$ denotes EOM for each field $\varphi^i$ and 
$\Theta(\varphi,\delta \varphi)$ denotes the  total surface term after integration by parts. 
The symmetry of the action is defined by the invariance of the action under the specific 
variation of fields $\varphi \rightarrow \varphi +\delta_{\epsilon}\varphi$. The invariance of the 
action under the symmetry
can be written as 
\beq \delta_{\epsilon} I [\varphi] = \int d^d x\, \sqrt{-g} \nabla_{a}K^{a} (\varphi, 
\delta_{\epsilon}\varphi) =0\,.  \label{SymVar}\eeq

The standard Noether current is introduced as 
\beq \tilde{J}^{a} \equiv    \Theta^{a}(\varphi,  \delta_{\epsilon}\varphi) - K^{a}(\varphi,  \delta_{\epsilon}\varphi)  
\label{eq:onshellcurrent} \,. \eeq
By taking the generic variation as the symmetry in Eq.~(\ref{SymVar}), 
one can see that this current satisfies 
\beq 
 \nabla_{a} \tilde{J}^{a} = -  E_{i}(\varphi)\delta_{\epsilon} \varphi^{i} \label{eq:onshellconser}\,, 
\eeq
which tells us  the on-shell conservation of the Noether current.  For global symmetry 
this procedure leads to the well-defined conserved charges by integrating the current 
on the hypersurface. However, that is not the case for a local symmetry. The basic 
reason for the inadequacy of this procedure in local symmetries is the existence of 
the so-called Noether identities.  In terms of local 
symmetry variation parameter $\epsilon=\epsilon(x)$, this  identity can be written as the form of 
\beq  - E_{i}(\varphi) \delta_{\epsilon} \varphi^{i} = \nabla_{a}S^{a}(E(\varphi)
, \delta_{\epsilon}\varphi )\,, \eeq
where $S^{a}$ denotes   the on-shell vanishing  current.  Even for the 
the global symmetries, which may appear as the rigid  limit of corresponding local symmetries, 
the current $S^{a}$ can be introduced. However, for this case the corresponding Noether identities are 
somewhat trivial. The 
absence of gauge fields in the case of global symmetry tells us that 
there are  missing equations for gauge fields among the equations of motion, $E(\varphi)=0$. 
 Because of these missing equations,  $S^{a}$ for the  global 
symmetry does not need to vanish even at the  the on-shell level. 

Noether identities together with the Poincar\'{e} lemma enable us to express the Noether current 
in terms of the on-shell vanishing current $S^{a}$ and a certain arbitrary 
anti-symmetric second rank tensor $J^{ab}$, which will be  named  as the {\it Noether potential} \footnote{This tensor, 
in the canonical terminology, is called as the {\it superpotential}. However, we have chosen  the 
{\it potential} for this one since the same terminology is also used in the context of (fake) 
supersymmetry with the completely different meaning.} 
\beq 
\tilde{J}^{a} = S^{a} +\nabla_{b}J^{ab} \label{eq:onshellJS}\,. 
\eeq
This expression shows that the Noether current for local symmetry becomes 
on-shell vanishing  up to a certain ambiguity.  Thus, it is unclear how to define 
conserved charges for local symmetry generically in terms of currents. To define 
conserved charges under the inherent ambiguity of the Noether current, one needs to adopt a certain prescription for choosing the appropriate current or potential. 

% The best way one can do is to construct quasi-local conservation. 
%Even though $J^{ab}$ seems to be ambiguous, 
%it has been shown that there is a concrete  way to construct it by local fields in the Lagrangian~\cite{Iyer:1994ys}, 
%and that the black hole entropy can be obtained from $J^{ab}$ unambiguously. 
%In order to define conserved charges for a local symmetry, it  has been shown that the relevant 
%quantity  is the potential not the current \cite{Iyer:1994ys}.  
%At the level of the total conserved charges which can 
%be defined at the asymptotic infinity, there  are some ways to define conserved
%charges. 

One way to define conserved charges is to introduce the asymptotically 
conserved potential through the on-shell vanishing current by linearizing the fields 
and EOM on a given background.  Then, by removing the ambiguity from the potential, 
one can define the conserved charges  as their integrals  of the
{\it on-shell}  potential corresponding to the `asymptotic Killing vectors'.  
More mathematically,  the arbitrariness of the potential may be resolved in the language of  the characteristic cohomology  
with the appropriate identification of  asymptotic boundary behaviors \cite{Barnich:2001jy}.  
The well-established  Arnowitt-Deser-Misner (ADM) \cite{ Arnowitt:1962hi} and Abbott-Deser-Tekin (ADT) \cite{Abbott:1981ff}
methods can be understood in this category.   

There are also well known approaches \cite{Wald:1993nt, Iyer:1994ys, Brown:1992br} 
to define conserved charges  by the Noether potential without the linearization,  which belong to quasi-local category.  
%At the level of  {\it quasi-local} conserved currents and charges, there are some  well-known approaches 
%including  Lagrangian Noether method \cite{Wald:1993nt, Iyer:1994ys}, canonical Hamiltonian method and covariant 
%phase space method.  
For some specific cases, these quasi-local charges are shown to be consistent with the conserved charges under  
the linearized method by using the one parameter family of solutions when a certain integrability condition is satisfied~
\cite{Barnich:2003xg}. This indicate that the conserved charges may be defined even at the non-linear 
level by the appropriate choice of the Noether potential.       

While the conventional Noether current defined in eq.(\ref{eq:onshellcurrent}) is conserved {\it on-shell} only,  it is possible  to 
construct the current which would be conserved without using EOM, {\it i.e.}  {\it off-shell} current.  
Indeed,  by using the eq.(\ref{eq:onshellJS}), one can construct the corresponding {\it off-shell}
Noether current $J^{a}$, as
\beq
J^{a}\equiv \tilde{J}^{a}- S^a  =  \nabla_{a}J^{ab}  \label{eq:onoffshellrel}\,.
\eeq
%Since we are interested in the radial 
%behavior of the angular momentum of certain black holes from the asymptotic infinity to the 
%horizon of black holes,  we  would like to consider the quasi-local charges without linearization.   
%In the following,  we will consider Einstein gravity and a certain 
%higher curvature gravity named as NMG  and will show what is   the relevant potential 
%in our case. It turns out that the (generalized) Komar expression for quasi-local charges is 
%enough for our purpose and  it will be shown that it  is consistent  with the previous results for 
%total charges by the linearization method.   
It turns out that this current satisfies the off-shell conservation law.  Just like the case of the on-shell current, there exists a  definite prescription  to obtain the {\it off-shell} 
Noether current by the appropriate potential $J^{ab}$, directly from the Lagrangian of any diffeomorphism theory of gravity~\cite{Padmanabhan:2009vy}. 
Since our derivation of   the horizon entropy and angular momentum invariance 
uses the {\it off-shell}  Noether current and its potential, we shall now review this approach\footnote{It is also possible to
extend the following analysis for the more general theories containing derivatives of curvature tensor. However, for the sake of
simplicity we shall concentrate on the Lagrangians constructed from the metric and curvature tensor.}.  
%and turns out to be easy to handle to obtain 
%some results, especially in some classical 
%gravity theories. In the following section, we will describe how to construct this current concretely 
%for some specific diffeomorphism invariant theory.
%%%%%%%%%%%%%%%%%%%%%%%%%%%%%%%%%%%%%%%%%%%%%%
% \subsection{Modified Noether currents and their potentials}\label{CCP2}
%%%%%%%%%%%%%%%%%%%%%%%%%%%%%%%%%%%%%%%%%%%%%%

Consider a generally covariant Lagrangian built from metric and curvature tensor
\begin{equation}
 I = \frac{1}{16\pi G}\int d^{d}x \sqrt{-g} \Big[\CL(g_{ab},R_{abcd}) + \CL_{m}(g_{ab},\phi) \Big]\,.
\end{equation}
The variation of $\CL$ with respect to  $g^{ab}$ is given by
\begin{equation}
\delta({\CL\sqrt{-g}}) = \sqrt{-g}(\mathcal{G}_{ab}\delta g^{ab}+\nabla_{a}\CV^{a}) \label{eq:varig1} \,,
\end{equation}
where 
\begin{eqnarray}
\mathcal{G}_{ab}&=& P_{a}^{cde}R_{bcde}   -2\nabla^{c}
\nabla^{d}P_{abcd}  -\frac{1}{2}g_{ab}\CL \quad ; \qquad P^{abcd} \equiv  \frac{\partial \CL}{\partial R_{abcd}}\,,  \nonumber\\
\CV^{a}(\delta g)&=& 2[P^{cbad}\nabla_{b}\delta g_{dc} - \delta g_{dc}\nabla_{b}P^{cabd}]\,.
\end{eqnarray}
Here, $\mathcal{G}^{ab}$ is the  generalized Einstein tensor which  satisfies Bianchi identity.
Let us consider the variation in metric which is  induced by the diffeomorphism 
\begin{equation}
x^{a}\rightarrow  x^{a} -  \xi^{a}  ~~;  ~~ \delta g_{ab} = 2 \nabla_{(a}\xi_{b)}
\label{eq:diff}\,. 
\end{equation}
Under the above transformation any scalar density changes as 
\begin{equation}
\delta_{\xi} (\CL \sqrt{-g}) =   \sqrt{-g}\nabla_{a}(\CL \xi^{a})\,.
\end{equation}
This implies  (using eq(\ref{eq:varig1})) the off-shell conservation law for the `modified'
Noether current
\bea
\nabla_{a}\lb 2\mathcal{G}^{ab}\xi_{b} + \xi^{a} \CL  -  \CV^{a}(\xi)\rb = \nabla_{a}J^{a} = 0\,, \label{eq:offconsr}
\eea
where 
\begin{equation}
J^{a} = 2\mathcal{G}^{ab}\xi_{b} + \xi^{a} \CL  -  \CV^{a}(\xi)\,. \label{ModNoe}
\end{equation}
 Expressing the boundary term $\CV^{a}$ as the linear combination of 
$\delta g_{ab}$ and $\delta \Gamma^{a}_{bc}$ \cite{Deruelle:2003ps} and using the eq. (\ref{eq:diff}),
we can rewrite the above expression as
 \begin{equation}
J^{a} = 2\nabla_{b}(P^{adcb} + P^{acdb})\nabla_{c}\xi_{d}+ 2P^{abcd}\nabla_{b}
\nabla_{c}\xi_{d} - 4\xi_{d}\nabla_{b}\nabla_{c}P^{abcd} \,. \label{eq:current1}
\end{equation}
There is an important point which we would like  to emphasize. The usual expression for
the Noether current consists of the last two terms of eq.(\ref{ModNoe}). A comparison with  
eq.(\ref{eq:onoffshellrel}) clearly shows that $S^{a}=2\mathcal{G}^{ab}\xi_{b}$. 
Consequently, this current is conserved only after using EOM. In contrast, the modified Noether current
obtained above obeys the {\it off-shell} conservation law. 
 
Next, we introduce the antisymmetric tensor $J^{ab}$ (Noether potential) by the condition
\begin{equation}
J^{a}= \nabla_{b}J^{ab}~.\label{eq:J1} 
\end{equation}
Then, we can take
\begin{equation}
J^{ab} = 2P^{abcd}\nabla_{c}\xi_{d} - 4\xi_{d}(\nabla_{c}P^{abcd})\,. \label{eq:Ncurrent}
\end{equation}  
It is worth mentioning that for Einstein gravity if the diffeomophisms are Killing vectors then the Noether potential reduces to the well known 
Komar potential \cite{Komar:1958wp}. As we shall show later, for the higher curvature gravity like NMG, the above Noether potential can also 
be used to calculate
the conserved quantities corresponding to the appropriate Killing vectors. In this sense, we  call  $J^{ab}[\xi_{Killing}]$ as the
generalized Komar potential.   

The conserved Noether charge corresponding to $J^{ab}$ may be defined by 
\begin{equation}
Q  =  \frac{1}{16\pi G}\int d\Sigma_{ab}\, J^{ab}. \label{eq:Q1}
\end{equation}
Our convention for the area element for the  subspace of codimension one and two is taken as $d\Sigma_a$ and $d\Sigma_{ab}$, respectively. 
These  are defined such that the Stokes'  theorem takes the following form
\[ \int_{\CM} d\Sigma_{a}\, J^{a} = \int _{\CM}d\Sigma_{a}\,  \nabla_{b}J^{ab} = \int_{\p \CM} d\Sigma_{ab}\, J^{ab}\,. \]

%where 
%\begin{equation}
%Q^{ab} \equiv \sqrt{-g}J^{ab}\,, \qquad P^{\mu} \equiv \sqrt{-g} {\cal J}^{\mu}\,.
%\end{equation}
%
The variations of the current  $J^{a}$ and the measure $d\Sigma_{a}$ under an  arbitrary 
diffeomorphism $x^{a} \rightarrow x^{a} - \xi'^{a}$ are given by
\beq
\delta_{\xi'}J^{a} = \xi'^{b}\nabla_{b}J^{a} - J^{b}\nabla_{b}\xi'^{a}\,, \qquad \delta_{\xi'}\, d\Sigma_{a} = d\Sigma_{a} (\nabla_{b}\xi'^{b})\,, \label{eq:variJ1}\ \eeq
which lead to the variation of the charge as 
\beq
\delta_{\xi'} Q =    \frac{1}{8\pi G}\int d\Sigma_{ab}\,  \xi'^{[b}J^{a]}\,. \label{eq:variQ1}
\eeq

One can apply the above  formalism even to gravity theory coupled with matters. To be specific, let us consider  an interacting scalar field with two derivatives, 
whose Lagrangian is 
\beq    \CL_{m} (g_{ab}, \phi) =  -\half \p_{a}\phi \p^{a}\phi - V(\phi) \,. \eeq
Its generic variation and diffeomorphism transformation are given by
\bea  \delta(\sqrt{-g}\CL_m) &=& \sqrt{-g}\Big[\CE_{\phi}\delta \phi +\nabla_{a}\CV^{a}(\delta \phi) \Big] \,,   \nn \\
        \delta_{\xi} (\sqrt{-g} \CL_{m}) & =&  \sqrt{-g} ( \xi^{a}\CL_{m}) \,, \eea
where $\delta_{\xi}\phi$ denotes the variation of the scalar field under diffeomorphism and  so it is given by $\delta_{\xi}\phi = \CL_{\xi}\phi = \xi^{a}\nabla_{a}\phi$.
The additional contribution from a scalar field can be computed as \footnote{In presence of  the matter term the right hand side of eq.(\ref{eq:varig1}) should contain an extra piece
proportional to $T_{ab}\delta g^{ab}$\,.}
\beq -2T^{ab}\xi_{b}  +  \xi^{a}\CL_{m}    -\CV^{a}(\delta_{\xi} \phi) =0\,. \eeq
As a result, the final expression for currents with a scalar field can be taken as the same form 
given earlier (\ref{ModNoe}) without matter fields.

%%%%%%%%%%%%%%%%%%%%%%%%%%%%%%%%%%%%%%%%%%%%%%%%%%%%
\section{Killing horizon and entropy of black holes}
In this section we shall compute the central extension  term in the Virasoro algebra among 
the generators corresponding to the diffeomorphism symmetry for the NMG
with a cosmological constant. Then, we implement  the stretched horizon approach
and calculate the horizon entropy for the non-extremal rotating BTZ black hole.  

Let us consider the action for  NMG with the cosmological constant  \cite{Bergshoeff:2009hq}
\beq I = \frac{1}{16\pi G}\int d^3x \sqrt{-g}\Big[ R + \frac{2}{\ell^{2}}
+ \frac{1}{m^2}\CK  \Big]\,, \label{eq:NMG} 
 \eeq
where
\[ 
\CK \equiv R_{ab}R^{ab} - \frac{3}{8}R^2\,. 
\]
For this Lagrangian the corresponding $P^{abcd}$  is given by
\beq P^{abcd} = \frac{\partial\CL }{\partial R_{abcd} }= g^{a [c}g^{d]b} + \frac{1}{m^2}
\Big(R^{a[c}g^{d]b} - R^{b[c}g^{d]a} - \frac{3}{4}
R  g^{a [c}g^{d] b}\Big)\,. \label{eq:P-tensor}
\eeq  
It is easy to see that  $\nabla_{a}P^{abcd} \ne 0$ in general and therefore 
we should use the general expressions for the Noether current and  potential given in 
(\ref{eq:current1}) and (\ref{eq:Ncurrent}), respectively. It is useful to compare this with the corresponding 
expressions in the usual Einstein or Lanczos-Lovelock gravity. In these cases only
second term in eq. (\ref{eq:current1}) and first term in eq.(\ref{eq:Ncurrent}) survive. 

Our task is to compute the central extension term $C[\xi_{1},\xi_{2}]$ for the 
algebra among the conserved Noether charges defined in eq.(\ref{eq:Q1}). The general form of $C[\xi_{1},\xi_{2}]$
is given by 
\bea
C(\xi_{1},\xi_{2}) = [Q(\xi_{1}),Q(\xi_{2})] - Q(\,[\xi_{1},\xi_{2}]\, )\,,\label{eq:central}
\eea
where the {\it Lie} bracket $[\xi_{1} , \xi_{2}]$  is  defined by 
\beq
[\xi_{1},\xi_{2} ]^{a} \equiv  \xi^{b}_{1}\nabla_{b} \xi^{a}_{2} -  \xi^{b}_{2}\nabla_{b} \xi^{a}_{1}\,.
\label{eq:diffalg}
\eeq
The essential part in the computation of the central extension term is to consider the $Lie$ variation of the
covariantly conserved Noether current under the diffeomorphism $x\rightarrow x- \xi_{2}$. 
Note that the current $J^{a}[\xi_{1}]$ is the consequence of the invariance of the original action 
under the diffeomorphism labeled by $\xi_{1}$ (see eq.(\ref{ModNoe})). The Lie variation of this current, 
$\delta_{\xi_{2}}J^{a}[\xi_{1}]$ induces the corresponding variation in the conserved Noether charge as
\beq
\delta_{\xi_{2}} Q[\xi_{1}] =    \frac{1}{16\pi G}\int d\Sigma_{ab}\,  \ls \xi^{b}_{2}J^{a}[\xi_{1}] - 
\xi^{b}_{1}J^{a}[\xi_{2}] \rs\,. \label{eq:variQ1}
\eeq
This enable us to compute the bracket among the Noether charges \cite{Majhi:2011ws}  
\beq
[Q(\xi_{1}),Q(\xi_{2})]  \equiv  \delta_{\xi_{1}}Q( \xi_{2}) - \delta_{\xi_{2}}Q(\xi_{1}) = 
2 \int  d\Sigma_{ab} \Big(\xi_{2}^{[a}J_{1}^{b]} - \xi_{1}^{[a}J_{2}^{b]}\Big) \,. \label{eq:QQB1}
\eeq
%By using eq.(\ref{eq:variQ1}), we arrive at
%\beq 
%[Q(\xi_{1}),Q(\xi_{2})] = 2 \int  d\Sigma_{ab} \Big(\xi_{2}^{[a}J_{1}^{b]} - \xi_{1}^{[a}J_{2}^{b]}\Big) \,
%\label{eq:QQB1}
%\eeq 
where $J^{a}_{1}\equiv J^{a}[\xi_{1}]$. Note that the the definition for the bracket is general in the sense that it does not require explicit form 
for the Noether current and works well for any covariant theory of gravity. 
% we consider the the following definition for the Lie bracket among the two diffeomorphisms
%\beq
%[\xi_{1},\xi_{2} ]^{a} =  \xi^{b}_{1}\nabla_{b} \xi^{a}_{2} -  \xi^{b}_{2}\nabla_{b} \xi^{a}_{1}\,,
%\label{eq:diffalg}
%\eeq
%\noindent Then the central extension $C[\xi_{1},\xi_{2}]$ to the Virasoro algebra  is given by 
%\bea
%C(\xi_{1},\xi_{2}) = [Q(\xi_{1}),Q(\xi_{2})] - Q(\,[\xi_{1},\xi_{2}]\, )\label{eq:central}
%\eea
%where, $[\xi_{1} , \xi_{2}]$  is  given by 
%\beq
%[\xi_{1},\xi_{2} ]^{a} =  \xi^{b}_{1}\nabla_{b} \xi^{a}_{2} -  \xi^{b}_{2}\nabla_{b} \xi^{a}_{1}\,,
%\label{eq:diffalg}
%\eeq
%By subsitutung eqs.(\ref{eq:current1}) and (\ref{eq:Ncurrent}) we get,
%\bea
%C(\xi_{1},\xi_{2}) &=& \frac{1}{8\pi G} \int d\Sigma_{ab} \lc\lb
%\ls 2 P^{bcde}\xi_{2}^{a}\nabla_{c}\nabla_{d}\xi_{1 e} + 
%\xi_{2}^{a} \ls 2 \nabla_{c}P^{bcde} \nabla_{d}\xi_{1e} - 4 \nabla_{d}P^{bcde}
%\nabla_{c}\xi_{1e}\rs \right.\right.\right.    \label{CentralTerm} \\
%&& \left.\left.\left. - 4 \xi_{2}^{a}\xi_{1e}\nabla_{c}\nabla_{d}P^{bcde}\rs - (1 \leftrightarrow 2)\rb - \frac{1}{2}\lb 2P^{abcd}\nabla_{c}\lc\xi_{1},\xi_{2}\rc_{d}
%- 4 \lc\xi_{1},\xi_{2}\rc_{d}\nabla_{c}P^{abcd} \rb
%\rc\,. 
%\nonumber ~~~~~
%\eea

Next, in order to obtain the horizon entropy we shall evaluate the algebra among the Noether 
charges, $Q([\xi_{1},\xi_{2}])$  on the null surface.  
To this purpose we implement  the stretched horizon approach given 
in \cite{Carlip:1999cy}. In this approach one begins with an
approximate Killing vector $\chi^{a}$ in the neighborhood of the boundary $\Sigma$ of generic 
$d+1$ dimensional Riemannian manifold. The local Killing horizon is defined by the condition 
$\chi^{2}=0$. This can be alternatively stated as
 \begin{equation}
\chi^{2} = \ep. \label{eq:B1}
\end{equation}
with $\ep$ being taken to be zero at the end. The vector orthogonal to the orbits of $\chi^{a}$ is
given by
\beq
\rho_{a} = -\frac{1}{2\kappa}\nabla_{a}\chi^{2}~. \label{eq:B2} 
\eeq
Where $\kappa$ is the surface gravity associated with the Killing vector field $\chi^{a}$
\beq
\kappa^{2} \equiv \lim_{\chi^{2}\rightarrow 0 } \lb\frac{\nabla_{a}\chi^{2} \nabla^{a}\chi^{2}}{4\chi^{2}}\rb\,.
\eeq
Consider a general diffeomorphism transformation 
\beq
\xi^{a} = T \chi^{a} + R \rho^{a} \label{eq:gendiff}\,,
\eeq
with the condition that
\beq
\delta \chi^{2} = 0 \qquad ; \qquad  as  \qquad \chi^{2} \rightarrow 0\,. \label{eq:condonchi}
\eeq
This condition leads to the following relation between two arbitrary functions $T$ and $R$:
\beq
R = \frac{1}{\kappa}\frac{\chi^{2}}{\rho^{2}} D T  \quad ;  \quad  D \equiv \chi^{a}\nabla_{a}\,.
 \label{eq:TRrel}  
\eeq
Now we compute the Noether potential $J^{ab}$ for the diffeomorphism (\ref{eq:gendiff}) on the 
stretched horizon.  By exploiting the relation (\ref{eq:TRrel}) we can express this diffeomorphism 
completely in terms of $T$. Then by  using eq.(\ref{eq:B10}) 
we get 
\bea
J^{ab}(\xi) = -  \lc P^{abcd}S_{cd} \lb 2\kappa T - \frac{D^{2}T}{\kappa}
\rb + 4\nabla_{c}P^{abcd} \ls\chi_{d}T +  \rho_{d}\frac{DT}{\kappa}\rs \rc \,.\label{eq:Jab1}
\eea 
Integrating this expression over the null surface with the differential area element $d\Sigma_{ab}$ given by (\ref{eq:nullsurfaceelement}) and using the identity (\ref{eq:Pidentity1}),
%\beq
%P^{abcd}\chi_{c}\rho_{d} =  -\frac{2|\chi|}{\rho \chi^{2}}P^{abcd}S_{cd} \label{eq:Pidentity1}
%\eeq
we arrive at the expression for the conserved Noether charge 
%First, we compute the Noether charge (\ref{eq:Q1}) 
%for NMG (\ref{eq:NMG}) under the diffeomorphism $\xi^{a}$. Using the 
%eqs. (\ref{eq:B3}, \ref{eq:B6}, \ref{eq:B10}) and using the identity
%and hence the expression for conserved Noether charge (\ref{eq:Q1}) becomes 
\bea
Q[\xi] =- \frac{1}{32\pi G} \int \sqrt{h}d^{d-2}x \lc P^{abcd}S_{ab}S_{cd} \lb 
2\kappa T - \frac{D^{2}T}{\kappa}\rb + 4\nabla_{c}P^{abcd} S_{ab}\ls\chi_{d}T - \rho_{d} 
\frac{DT}{\kappa}\rs \rc\,.  
\eea 
Note that the above expression contains the terms proportional to $P^{abcd}$ as well as $\nabla_{a}P^{a\cdots}$. 
However, near the Killing horizon,  $\nabla_{a}P^{a\cdots}$ term is of the higher order $\chi^{2}$ compared to  
$P^{abcd}$ term. Since the Killing horizon is defined by the limit $\chi^{2}\rightarrow 0$,  we can neglect 
the derivative term of $P$-tensor. \footnote{For some specific cases,  where the near horizon geometry is $AdS$ ({\it e.g} BTZ black hole) or the 
product of two maximally symmetric spaces like $AdS_{2}\times {\cal M}^{n-2}$ or $Rindler \times {\cal M}^{n-2}$, 
 $\nabla P$ vanishes identically.} 
This can be explicitly checked by Taylor expanding the $P$ and $\nabla P$ terms near 
the Killing horizon. Similar kind of arguments were used earlier (see the eq.(19) of Ref. \cite{Cvitan:2004pd}) 
to obtain the central charge for higher curvature gravity within the Carlip's on-shell approach. 
On using this fact, the expression for the Noether charge becomes 
\bea
Q[\xi] =- \frac{1}{32\pi G} \int \sqrt{h}\, d^{d-2}x \lc P^{abcd}S_{ab}S_{cd} \lb 
2\kappa T - \frac{D^{2}T}{\kappa}\rb  \rc\,.  \label{eq:Q2}  
\eea 
%This is the most general expression for the conserved Noether charge corresponding to the diffeomorphism (\ref{eq:gendiff}). 
%For the NMG Lagrangian (\ref{eq:NMG}) the
%expression for $P^{abcd}$ ( see \ref{eq:P-tensor}) depends on $g^{ab}, R$ and $R^{ab}$.
%The AdS black hole geometries which we are interested in this work  satisfy $R^{ab} 
%\sim g^{ab}$ and consequently, the terms proportional to the divergence of $P$-tensor give
%no contribution. For non-AdS geometries however, the near horizon geometry becomes 
%$Rindler \times {\cal M}^{n-2}$ for the non-extreamal black holes, while  $AdS_{2}\times 
%{\cal M}^{n-2}$ for the extreamal black holes. In these cases by Taylor expanding the
%$\nabla P$ terms near the horizon we can easily see that they give vanishing contribution. 
%Similar kind of arguments were used in [ref(Cvitan et al.)] to obtain the central charge 
%for higher curvature gravity within the Carlip's on-shell approach [carlip-1]. In view of this 
%we can write the conserved Noether charge (\ref{eq:Q2}) as
% \bea
%Q[\xi] =- \frac{1}{32\pi G} \int \sqrt{h}d^{d-2}X \lc P^{abcd}S_{ab}S_{cd} \lb 
%2\kappa T - \frac{D^{2}T}{\kappa}\rb +  \rc ~. \label{eq:Q2}  
%\eea 

We now evaluate the corresponding expression for $Q(\, [\xi_{1},\xi_{2}]\, )$. First, we note that
the Lie bracket for the diffeomorphisms (\ref{eq:gendiff}) can be writen as 
\bea
[\xi_{1},\xi_{2}]^{a} = \chi^{a}[T_{1},T_{2}] - \frac{\rho^{a}}{\kappa}D[T_{1},T_{2}] \quad  ; \quad
[T_{1},T_{2}] =  T_{1}DT_{2}- T_{2}DT_{1} \,.   \label{eq:lbxiT}
\eea
%with
%\beq
%[T_{1},T_{2}] = \lb T_{1}DT_{2}- T_{2}DT_{1}\rb \,. \label{eq:lbT}
%\eeq
%Use of eqs.(\ref{eq:B4},\ref{eq:B6}) in place of the diffeomorphisms in (\ref{eq:diffalg}), yields 
%\bea
% [T_{1}, T_{2}] &=&  T_{1}DT_{2} -  T_{2}DT_{1}\,,    \\
%D[ T_{1}, T_{2}] &=&  T_{1}D^{2}T_{2} -  T_{1}D^{2}T_{2}\,,   \\
%D^{2} [ T_{1}, T_{2}] &=&  \Big( DT_{1} D^{2}T_{2} + T_{1}D^{3}T_{2}\Big)  -  
%(1\leftrightarrow 2) \,. 
%\eea
Inserting this in eq.(\ref{eq:Q2}), we obtain the expression for $Q(\, [\xi_{1},\xi_{2}]\, )$ as
%\bea
%\nabla_{c}P^{abcd} S_{ab}\chi_{d} = -\frac{2|\chi|}{\rho \chi^{2}}\nabla_{c}P^{abcd}\chi_{a}\chi_{d}\rho_{b}
% \  \   ; \ \ \nabla_{c}P^{abcd} S_{ab} \rho_{d} = \frac{2|\chi|}{\rho \chi^{2}}\nabla_{c}P^{abcd}\rho_{a}\rho_{d}\chi_{b}
%\eea
%\bea
%Q(\, [\xi_{1},\xi_{2}]\, ) &=& - \frac{1}{32\pi G} \int d^{d-2}x\sqrt{h}  \lc  P^{abcd}
%S_{ab}S_{cd} \lb 
%2\kappa \ls T_{1}DT_{2} - T_{2}DT_{1}\rs -\frac{1}{\kappa} \ls DT_{1} D^{2}T_{2} 
%\right.\right. \right.\nonumber \\  
%&& \left.\left.\left. 
%+ T_{1}D^{3}T_{2}  - DT_{2} D^{2}T_{1} - T_{2}D^{3}T_{1}\rs  \frac{}{}\rb
%- \frac{8}{\chi^{2}}\nabla_{c}P^{abcd}\lb \frac{}{} 
%\chi_{a}\rho_{b}\chi_{d}(T_{1}DT_{2}-T_{2}DT_{2})
%  \right.\right.  \nn \\
%&& \left.\left. \frac{}{}
%- \frac{1}{\kappa}\rho_{a}\rho_{d}\chi_{b}\times (T_{2}D^{2}T_{1}-T_{1}D^{2}T_{2})\rb 
% \rc ~. \label{eq:Q3}  
%\eea 
 \bea
Q[\lc\xi_{1},\xi_{2}\rc] &=& - \frac{1}{32\pi G} \int \sqrt{h}\, d^{d-2}X  \lc  P^{abcd}
S_{ab}S_{cd} \lb 
2\kappa \ls T_{1}DT_{2} - T_{2}DT_{1}\rs -\frac{1}{\kappa} \ls DT_{1} D^{2}T_{2}
 \right.\right. \right. \nonumber \\  
&& \left.\left.\left.
+ T_{1}D^{3}T_{2}  - DT_{2} D^{2}T_{1} - T_{2}D^{3}T_{1}\rs  \frac{}{}\rb
 \rc ~. \label{eq:Q3}  
\eea 
 
\noindent Next, we perform the similar analysis on the right hand side of eq.(\ref{eq:QQB1}). The explicit expression of
the Noether current (\ref{eq:current1}) for the diffeomorphism (\ref{eq:gendiff}) is given by
%by using eq.(\ref{eq:J1}) and eq.(\ref{eq:Ncurrent}) we get
%\bea
%J^{a} (\xi) &=& 2 P^{abcd} \nabla_{b}\nabla_{c}\xi_{d}+
%2 \nabla_{b}P^{abcd} \nabla_{c}\xi_{d} - 4 \nabla_{b}\ls \xi_{d} \nabla_{c}P^{abcd}\rs
% \label{eq:J2}
%\eea
\bea
J^{a}(\xi) &=& 2 P^{abcd} \ \frac{\chi_{c}\rho_{d}\chi_{b}}
{\chi^{4}}\ls 2\kappa DT - \frac{D^{3}T}{\kappa}\rs + 2 \nabla_{b}P^{abcd} 
 \ \frac{\chi_{c}\rho_{d}}{\chi^{2}}
\ls2\kappa T - \frac{D^{2}T}{\kappa}\rs   \nonumber\\
&&  - 4 \nabla_{b}\lb \ls T\chi_{d}-\frac{DT}{\kappa}\rho_{d}\rs \nabla_{c}P^{abcd}\rb \,.
 \label{eq:J3}
\eea
Substituting this in eq.(\ref{eq:QQB1}) and using the identity (\ref{eq:Pidentity2}), we can see
\bea
[Q(\xi_{1}),Q(\xi_{2})] &=& \frac{1}{32 \pi G} \int \sqrt{h}\, d^{d-2}x  \lc 
P^{abcd}S_{ab}S_{cd}\lb\ls2\kappa DT_{1} - \frac{D^{3}T_{1}}{\kappa} \rs T_{2}- 
(1 \leftrightarrow 2)\rb\rc. \label{eq:Q1Q2}
\eea
As before, we have neglected the terms proportional to $\nabla P$ and $\nabla\nabla P$.  
Combining this result with the eq.(\ref{eq:Q3}), the central extension term $C[\xi_{1},\xi_{2}]$ finally becomes
\bea
 C[\xi_{1},\xi_{2}] &=& -\frac{1}{32\pi G} \int \sqrt{h}\, d^{d-2}X \lc 
P^{abcd}S_{ab}S_{cd}\frac{1}{\kappa} \lb DT_{1}D^{2}T_{2} -DT_{2}D^{2}T_{1} \rb  \rc\,. \label{eq:C1}
\eea
%\bea
% C(\xi_{1},\xi_{2}) &=& -\frac{1}{32\pi G} \int \sqrt{h}d^{D-2}x \lc 
%P^{abcd}S_{ab}S_{cd}\frac{1}{\kappa} \lb DT_{1}D^{2}T_{2} -DT_{2}D^{2}T_{1} \rb
%\right. \nonumber\\
%&& \left. +2 \nabla_{b}P^{abcd} S_{cd}\lb \ls T_{2}\rho_{a}- \frac{DT_{2}}{\kappa}\chi_{a}\rs
%\ls2\kappa T_{1} - \frac{D^{2}T_{1}}{\kappa} \rs - (1 \leftrightarrow 2)\rb 
%\right. \label{eq:C1}\\
%&& \left. + 8 \nabla_{b}\nabla_{c} P^{abcd}  \lb \ls T_{1}\chi_{d}-\frac{DT_{1}}{\kappa}\rho_{d}\rs
%\ls T_{2}\rho_{a}-\frac{DT_{2}}{\kappa}\chi_{a}\rs - (1 \leftrightarrow 2) \rb
%\rc \nonumber
%\eea
%The AdS black hole geometries which we are interested in this work  satisfy $R^{ab} 
%\sim g^{ab}$ and consequently, the $\nabla P, \ \nabla\nabla P$ terms give
%no contribution. For non-AdS geometries however, the near horizon geometry becomes 
%$Rindler \times {\cal M}^{n-2}$ for the non-extreamal black holes, while  $AdS_{2}\times 
%{\cal M}^{n-2}$ for the extreamal black holes. In these cases, by Taylor expanding the
%$\nabla P$ terms near the horizon we can easily see that they give vanishing contribution. 
%Similar kind of arguments were used in \cite{Cvitan:2004pd} to obtain the central charge 
%for higher curvature gravity within the Carlip's on-shell approach \cite{Carlip:1999cy}. In view of this 
%we can write the central term as
%\bea
% C (\xi_{1},\xi_{2}) &=& -\frac{1}{32\pi G} \int \sqrt{h}d^{D-2}x \lc 
%P^{abcd}S_{ab}S_{cd}\frac{1}{\kappa} \lb DT_{1}D^{2}T_{2} -DT_{2}D^{2}T_{1} \rb  \rc \label{eq:C2}
%\eea
By rewriting the above expression in the Fourier variables 
\bea
 C(\xi_{1},\xi_{2}) &=&   \sum_{mn} f_{mn} C(\xi_{m},\xi_{n})\,, \label{eq:C2} 
\eea
we obtain
\beq
C(\xi_{m},\xi_{n}) =  -\frac{1}{32\pi G} \int \sqrt{h}\, d^{d-2}x \lc 
P^{abcd}S_{ab}S_{cd}\frac{1}{\kappa} \lb DT_{m}D^{2}T_{n} -DT_{n}D^{2}T_{m}\rb\rc \,.
\label{eq:fourierC}
\eeq
We now apply this analysis for non-extremal  rotating BTZ black hole given by the metric
 \beq 
 ds^2 = L^2\bigg[- \frac{(r^2-r^2_{+})(r^2- r^2_{-})}{r^2} dt^2  +  
\frac{r^2} {(r^2-r^2_{+})(r^2- r^2_{-})}dr^2 + r^2\Big(d\theta - 
\frac{r_{+}r_{-}}{r^2}dt\Big)^2\bigg]\,.\label{eq:BTZ1}
\eeq
For this case,  the approximate Killing vector $\chi^{a}$  is given by
\bea
\chi^{a} = (1,0, \O) \label{eq:BTZKillingupp} \quad  ;  \quad  
\chi_{a} = \ls g_{tt} + \O g_{t\theta}, 0 , \O g_{\theta\theta}+g_{t\theta} \rs\,, \label{eq:BTZKillinglow}
\eea
where $\O = r_{-}/r_{+}$ is the angular velocity at the outer horizon.
In the near horizon region given by $r = r_{+} + \ep$, it is easy to see that the norm of the
Killing vector $\chi_{a}$ behaves as
\beq
\chi^{2} = -\ep \lb\frac{2(r_{+}^{2}-r_{-}^{2})}{r_{+}} L^{2}\rb\,. \label{eq:BTZnormKilling}
\eeq

In order to obtain the central charge, we take the ansatz of $T_{n}$, appropriate for Diff$(S^{1})$ algebra \cite{Silva:2002jq}, as 
\beq
T_{n} = \frac{1}{\alpha + \O } \exp[{i n (\alpha t + \theta + g(r))}]~. \label{eq:T}
\eeq 
Here, variables $t$ and $\theta$ have periodicities $2\pi/\alpha$ and $2\pi$ respectively and 
$g(r)$ is a certain function regular at the Killing horizon. Inserting this in eq.(\ref{eq:fourierC}), we arrive at
\bea
C[\xi_{m},\xi_{n}] &=&   \frac{\cal A}{8\pi G} \lb-i m^{3} \delta_{m+n,0}\, \frac{(\alpha + \O)}{\kappa}\rb\,,
\label{eq:fourierC1}
\eea
where 
\beq
{\cal A} \equiv -\frac{L}{2} \int r d\theta  P^{abcd}S_{ab}S_{cd}\,. \label{eq:A}
\eeq

Using the anstaz of $T_{n}$, one can see that the Fourier modes of the Noether charges are given by 
\bea
Q (\xi_{m}) = \frac{\cal A}{8\pi G} \lb \delta_{m,0}\frac{\kappa}{(\alpha + \O)}\rb \quad ; 
\quad Q(\, [\xi_{m},\xi_{n} \,] ) = -i (m-n)Q (\xi_{m})\,. \label{eq:fourrierQ}
\eea
Using these expressions and the eq.(\ref{eq:fourierC1}) in the algebra (\ref{eq:central}), we identify
the central charge $c$ and the zero-mode eigenvalue of $Q(\xi_{m}) $ as
\bea
c = \frac{\alpha+\O}{\kappa}\frac{12}{8 \pi G}{\cal A} \quad ; 
\quad Q_{0} = \frac{\kappa}{(\alpha + \O)}\frac{\cal A}{8\pi G} \label{eq:centralcharge1}\,.
\eea
Through the Cardy formula, the entropy for non-extremal BTZ black holes is given by
\bea
S = 2\pi \sqrt \frac{c\Delta_{0}}{6} = 2\pi \sqrt \frac{c\ls Q_{0}-\frac{c}{24}\rs}{6} = 
\frac{{\cal A}}{4G}\lb2-\ls\frac{\alpha+\O}{\kappa}\rs^2\rb^{1/2}\,.
\eea
By setting $\alpha= \kappa-\O$, one can see that the above expression reduces to familar Wald formula \cite{Wald:1993nt}. 
 
Using the $P$-tensor for the rotating BTZ black holes
\beq
P^{abcd} = g^{a[c}g^{d]b}\lb1+\frac{1}{2m^{2}L^{2}}\rb\,, \label{eq:P-tensorBTZ2}
\eeq	
and the expression for $S_{ab}$ in the eq.(\ref{eq:A}), we obtain
\beq
{\cal A} = 2\pi L r_{+}\lb1+\frac{1}{2m^{2}L^{2}}\rb\,. \label{eq:ABTZ}
\eeq
Hence, the corresponding entropy of the horizon (with the choice $\alpha= \kappa-\O$) becomes
\beq
S = \frac{\pi L r_{+} }{2G}\lb1+\frac{1}{2m^{2}L^{2}}\rb~. \label{eq:entropyBTZ}
\eeq
The first term in the above represents the entropy for rotating BTZ black hole in Einstein gravity, while the 
second term gives its correction due to the higher curvature terms in the Lagrangian (\ref{eq:NMG}). 
Our result for entropy is also consistent with the one given in \cite{Saida:1999ec}.
%This is the generalisation of eq. ($28$) in [ref(paddy-bibhas-1)] for the case where $\nabla_{a}
%P^{abcd} \ne 0$. However, by Taylor expanding last two terms near the horizon we can easily
%see that the relative order difference between these two terms and the first one is $O(\chi^{2})$.
%Consequently, we can neglect the terms  proportional to divergence of $P^{abcd}$ \footnote{Similar 
%arguments were used by [ref(Cvitan et al.)] where they geeralised Carlip's on shell approach for 
% higher curvature gravity.}.  For the AdS geometries which we are interested in the terms proportional 
%to divergence of $P^{abcd}$ vanish indentically and we left with only the first term in eq.(\ref{eq:C}) 
%[we have to write this para more carefully]. 
  
%\noindent{\it Comment on extremal BTZ:}\\
We now briefly describe some important steps to compute the entropy for extremally rotating BTZ black holes 
by the {\it off-shell} expressions for Noether current and potential. 
For the extremal case, the `stretched horizon' approach, given in \cite{Carlip:1999cy} needs
to be modified. An alternate way to deal with such geometries within the stretched 
horizon approach was presented in \cite{Carlip:2011ax}. We shall not follow this approach here. Instead, 
we consider the near horizon limit of extremally rotating BTZ black hole in which case  the resultant geometry is $AdS_{3}$. 
We then compute the central extension term for the diffeomorphisms which preserve the asymptotic fall-off conditions \cite{Brown:1986nw, Guica:2008mu}. 

%The central charge 
% 
%The Noether current and potential 
%is now computed  
%We then compute the central extension term (\ref{eq:eq:central}) on the asymptotic boundary of this near horizon extremal geometry. Unlike the stretched horizon approach in this case we consider diffeomorphism
%are chosen 
%
%the usual  analysis of \cite{Brown:1986nw, Guica:2008mu} and calculate the central charge on the asymptotic infinity of near horizon 
%extremal geometry. 
%
%we briefly 
%mention few steps to compute the entropy of extremal BTZ black hole using the usal 
%analysis given in \cite{Brown:1986nw, Guica:2008mu}.
 To begin with, let us take the near horizon extremal BTZ geometry in the form of 
\beq  ds^2_{NH} = \frac{L^2}{4}\bigg[ -\rho^2dt^2  + \frac{1}{\rho^2}d\rho^2 
+4r^2_H\Big(d\theta - \frac{\rho}{2r_H}dt\Big)^2\bigg]\,.  \eeq

\noindent The relevant diffeomorphisms given in \cite{Compere:2012jk,Guica:2008mu}, are
\beq \xi_{n} = e^{in\theta}\Big(\p_{\theta} -in\rho \p_{\rho}\Big)\,, \eeq
which preserve  appropriate boundary conditions  \cite{Brown:1986nw}.
It can be easily checked that these diffeomorphisms satisfy Diff$(S^{1})$ algebra (\ref{eq:Diffalgebrafourier}).
We now use  the generic expression for the central term 
\bea
C(\xi_{1},\xi_{2}) &=& \frac{1}{8\pi G} \int d\Sigma_{ab} \lc\lb
\ls 2 P^{bcde}\xi_{2}^{a}\nabla_{c}\nabla_{d}\xi_{1 e} + 
\xi_{2}^{a} \ls 2 \nabla_{c}P^{bcde} \nabla_{d}\xi_{1e} - 4 \nabla_{d}P^{bcde}
\nabla_{c}\xi_{1e}\rs \right.\right.\right.    \label{CentralTerm} \\
&& \left.\left.\left. - 4 \xi_{2}^{a}\xi_{1e}\nabla_{c}\nabla_{d}P^{bcde}\rs - (1 \leftrightarrow 2)\rb - \frac{1}{2}\lb 2P^{abcd}\nabla_{c}\lc\xi_{1},\xi_{2}\rc_{d}
- 4 \lc\xi_{1},\xi_{2}\rc_{d}\nabla_{c}P^{abcd} \rb
\rc\,. 
\nonumber ~~~~~
\eea
which can be obtained by using the eq.(\ref{eq:central}). It may be recalled that this central extension was derived 
using the off-shell expressions for Noether current (\ref{eq:current1}) and potential (\ref{eq:Ncurrent}). 
After performing a little algebra,  we obtain 
\beq  C(\xi_{p}, \xi_{q} ) =  -i \frac{L}{8G}\Big[ 1 + \frac{1}{2m^2L^2}  \Big] \, p(p^2+4r^2_H)\, \delta_{p+q,0} \,, \eeq
%
%
% =-i \frac{c_{IR}}{12}p(p^2+a)\delta_{p+q,0}
and hence the central charge becomes, 
\beq c_{IR} = \frac{3L}{2G}  \Big[ 1 + \frac{1}{2m^2L^2}  \Big] \,. \eeq
The above expression for  the central charge is also consistent with the corresponding one given in~\cite{Azeyanagi:2009wf}. 
By exploiting the standard Cardy formula we can obtain the  entropy for the  infrared dual CFT as
\beq
S_{IR}= 2\pi\sqrt{\frac{c_{IR}}{6}(M_{H} L+J_{H} ) } + 2\pi \sqrt{\frac{c_{IR}}{6}(M_{H} L - J_{H})} = 2\pi\sqrt{\frac{c_{IR}}{3}J_{H}}\,, \label{eq:entropyextremBTZ}
\eeq
where $M_{H}$ and $J_{H}$ represent the mass and angular momentum at the horizon, respectively.  
The second equality in the above expression comes from the fact that, 
at extremality, mass and angular momentum satisfy $M_{H}L=J_{H}$ . 
As we shall see later, the quasi-local angular momentum $J_{H}$ matches exactly 
with the total angular momentum $J_{\infty}$. This property is the consequence of the invariance 
of the angular momentum along the radial direction.  
In the next section we shall give the explicit proof for this invariance and show that the 
entropy (\ref{eq:entropyextremBTZ}) is indeed identical with the one computed earlier (\ref{eq:entropyBTZ}).
%%%%%%%%%%%%%%%%%%%%%%%%%%%%%%%%%%%%%%%%%%%%%%%%%%
\section{Angular momentum and extremal black holes}
%%%%%%%%%%%%%%%%%%%%%%%%%%%%%%%%%%%%%%%%%%%%%%%%%%
In this section we will compute the angular momentum for the BTZ black holes in NMG
by  using only the {\it off-shell} Noether potential (\ref{eq:Ncurrent}). As will be shown later,  the angular momentum remains 
invariant along the radial direction. Next, we show that  hairy deformation of the extremal BTZ black holes also carry the same property. 
The point is that these extremal black hole solutions interpolate two AdS spaces, {\it viz.} the asymptotic $AdS_3$ and the near horizon $AdS_3$ and 
therefore can be regarded as a holographic realization of RG flows of a certain field theory. 
Our confirmation of the angular momentum invariance in these black hole solutions also verifies that the entropy of the infrared dual CFT is 
identical with the BH entropy\footnote{The relevant discussion for the Einstein gravity is provided in Appendix-B.}. 

The action for  NMG minimally coupled to a scalar field $\phi$ is

\beq I = \frac{1}{16\pi G}\int d^3x \sqrt{-g}\Big[ R + 
\frac{1}{m^2}\CK          - 
\half \p_{a}\phi \p^{a}\phi - V(\phi) \Big]\,. \label{eq:NMGscal}
 \eeq
The equations of motion for the metric and the scalar field are given by
\beq \CE_{ab} \equiv \CG_{ab}  - T_{ab} =0\,, \qquad  \CE_{\phi}\equiv \nabla^2\phi - \p_{\phi}V =0\,,  \eeq
where $\CG_{ab}$ denotes the generalized Einstein tensor and $T_{ab}$ does the stress tensor for a scalar field as
\bea \CG_{ab} &=&  R_{ab} -\half  R g_{ab}   +{1 \over {2 m^2}} \CK_{ab} \,,  \qquad  T_{ab} =  \half \p_{a}\phi \p_{b}
\phi  -\half g_{ab} \Big[ \half\p_{c}\phi\p^{c}\phi +   V(\phi)\Big]\,, 
\\ \nn \\
\CK_{ab} &\equiv& 
g_{ab}\Big(3R_{cd}R^{cd}-\frac{13}{8}R^2\Big)
+ \frac{9}{2}RR_{ab} -8R_{ac}R^{c}_{b}+
\half\Big(4\CD^2R_{ab}-\CD_{a}\CD_{b}R
-g_{ab}\CD^2R\Big)\,.  \nn \eea

\noindent By applying the formalism given in section-\ref{CCP} to the present case, we obtain 
the quasi-local angular momentum  corresponding to the rotational Killing vector on the domain 
$\CB$ of codimension two, as
\beq J_{\CB} \equiv Q_{\CB}(\xi_R) = \frac{1}{16\pi G}\int_{\CB}  d\Sigma_{ab}\, J^{ab}
(\xi_R)   = \frac{1}{8G}\sqrt{-\det g}~ J^{rt}(\xi_R) \bigg|_{\CB}   \,,  \label{AngMom} \eeq
where $\det g$ denotes the determinant of the three-dimensional metric $g_{ab}$.  
Our convention for the normalization of the rotational Killing vector $\xi_R \equiv 
\frac{\p}{\p\theta}$  is  chosen such as ${\xi^2_R}_{(r\rightarrow \infty)}  \rightarrow L^2r^2$.

%In this case of NMG, the $P$-tensor  is given by
%%
%\beq P^{abcd} = g^{a [c}g^{d]b} + \frac{1}{m^2}
%\Big(R^{a[c}g^{d]b} - R^{b[c}g^{d]a} - \frac{3}{4}
%R  g^{a [c}g^{d] b}\Big)\,. \label{eq:P-tensor}
%\eeq  
%
For the  Killing vector $\xi^{\mu}$ of the metric,  one can see that
\bea
 J^{a} 
&=& 2R^{ab}\xi_{b} -\frac{2}{m^2}R^{cd}\nabla_{c}
\nabla_{d}\xi_{b}   \nn \\
&& + \frac{1}{m^2}\bigg[-2R^{acdb}R_{cd}-\frac{3}
{2}RR^{ab} +2\nabla^2R^{ab}-\frac{1}{2}\nabla^{a}
\nabla^{b} R- \frac{1}{2}g^{ab}\nabla^2R\bigg]\xi_{b}\,. \eea

\noindent Let us consider rotating BTZ black holes  (\ref{eq:BTZ1}) as an example to give some taste of  our approach. 
By computing the Komar potential of BTZ black holes, one obtains 
\beq J^{rt}(\xi_R)  =\frac{2r_{+}r_{-} }{ rL^2} \Big[1 + \frac{1}
{2m^2L^2}\Big] \,,  \eeq
which leads to the quasi-local  angular momentum as 
\beq J_{\CB_r} =  J_{\infty} = J_{H} = 
\frac{Lr_{+}r_{-} }{4G}\Big[1 + \frac{1}{2m^2L^2}\Big]\,,\eeq
where $J_{\CB_{r}}$, $J_{\infty}$ and $J_{H}$ denote the quasi-local angular momentum at the $r=constant$,  
the asymptotic infinity and the horizon, respectively.  Note that this expression for the angular momentum at asymptotic infinity, 
$J_{\infty}$ is  consistent with the result for the angular momentum of BTZ 
black holes in other methods like ADT or boundary stress tensor~\cite{Saida:1999ec,Clement:2009gq,Hohm:2010jc,Nam:2010ub,Devecioglu:2010sf,Ghodsi:2011ua}. In fact, 
this result shows us that  the angular momentum is invariant along the
 radial direction as was argued generically to be the case for pure Einstein 
gravity in~\cite{Julia:1998ys}. Now it is easy to see that by inserting  $J_{H}$ in the eq.(\ref{eq:entropyextremBTZ}) we can reproduce the entropy for extremally rotating BTZ black holes
\beq
S_{IR} =  \frac{\pi L r_{+} }{2G}\lb1+\frac{1}{2m^{2}L^{2}}\rb~. \label{eq:entropyextBTZ2}
\eeq
It is interesting to note that our result agrees with the corresponding one obtained  
by using the central charge of ultraviolate CFT ($ c_{UV}$) and the total angular momentum $J_{\infty}$  
\cite{Strominger:1997eq,Saida:1999ec}. 
 
Now we consider the extremally rotating BTZ black holes in NMG coupled with a scalar
field and show explicitly the invariance of the angular momentum along the radial direction.
%This would allow us realize non-trivial holographic RG flow for the dual field theory. 
The most generic axi-symmetric  metric of the rotating  hairy  black holes, 
deformed from BTZ ones,  can be taken as 
\beq ds^2 = L^2\Big[ -e^{2A(r)}dt^2 +e^{2B(r)}dr^2 + r^2(d\theta + e^{C(r)}dt)^2\Big]\,,
\label{eq:deformBTZ1} \eeq
with the asymptotic $AdS$ boundary condition for $r\rightarrow \infty$ as
\beq e^{A(r)} \rightarrow r \,, \qquad e^{B(r)} \rightarrow \frac{1}{r}\,, 
\qquad e^{C(r)}\rightarrow constant \,, \qquad \phi(r) \rightarrow constant\,.  \eeq
To obtain the extremal hairy AdS black hole solutions in the case of NMG, the scalar potential can be taken in terms of the so-called superpotential $\CW(\phi)$ as
\beq V(\phi) = \frac{1}{2L^2}(\p_{\phi}\CW)^2\Big[1-\frac{1}{8m^2L^2}
\CW^2\Big]^2 - \frac{1}{2L^2}\CW^2\Big[1-\frac{1}{16m^2L^2}\CW^2\Big]\,, \eeq
which is motivated by the domain wall case~\cite{Louzada:2010jq}
 and can be explained by fake supersymmetry in the Einstein gravity limit \cite{Hyun:2012bc}. 
As was shown in~\cite{Kwon:2012zh},  extremal hairy deformed BTZ black holes satisfy 
some reduced EOM.  By solving partially this reduced EOM, one can show that the metric functions 
$A$ and $B$ are determined  in terms of a certain function $\Psi$ and the 
superpotential $\CW$ as
\bea && rA'(r) -1 = \frac{\Psi}{r}e^{B(r)}\,,   \label{MFNMGa}\\ 
&&e^{-B(r)} = \frac{r}{2}\Big[\CW - \frac{\Psi}{r^2}\Big]\,, \label{MFNMG} \eea
where $'$ denotes the differentiation with respect to the radial coordinate $r$.
The remaining part of the  reduced EOM for the function $\Psi$ is given by
\beq \tilde{\Delta}\Big[1+ \frac{1}{2m^2L^2}\Big] = \frac{1}{m^2L^2}e^{-2B}
(\ddot{\Psi} -\CW\dot{\Psi}) + \Big[1+ \frac{1}{8m^2L^2}(\CW^2- 4\dot{\CW})
\Big]\Psi\,,   \eeq
where $\tilde{\Delta}$ is a certain integration constant and $\dot{}$ denotes the differentiation 
defined as $\dot{\Psi} \equiv e^{-B}\Psi'$.  In fact, it turns out that the constant 
$\tilde{\Delta}$ is related to the horizon values of $\CW$ as 
\beq \tilde{\Delta} = r^2_H\CW(\phi_H) \Big[1 + \frac{1}{8m^2L^2}\CW^2(\phi_H)
\Big] \Big[ 1 + \frac{1}{8m^2L^2}\CW^2(\phi_{\infty})\Big]^{-1}\,.    \label{PsiNMG}  \eeq
Note that these solutions satisfy the extremality condition as $e^{-2B(r_H)} =
\frac{d}{dr} e^{-2B(r_H)}=0$. This is equivalent to the relation between the mass and the angular momentum : $M_{\infty}L=J_{\infty}$.

In the following  we confine ourselves to the Brown-Henneaux boundary conditions~\cite{Brown:1986nw}. These boundary conditions however are not the most general ones even for the usual Einstein gravity coupled to a scalar field \cite{Henneaux:2002wm,  Martinez:2004nb}. The asymptotic analysis in this case gives us the following expression for metric variables, the superpotential and the scalar field respectively 
\bea && A(r) = \ln r - \frac{\tilde{\Delta}}{2r^2}+ \cdots\,, \qquad B(r) = -\ln r + \frac{1}
{2r^2}\Big(\tilde{\Delta}- \frac{1}{2q}\tilde{\phi}^2_1\Big) + \cdots\,,  \\ 
       \nn   \\  && \CW(\phi(r)) = 2 + \frac{\tilde{\phi}^2_1}{2qr^2} + \cdots\,, \qquad 
\phi(r) = \phi_{\infty} + \frac{\tilde{\phi}_1}{r} + \cdots\,, \nn\eea
where $q$ is defined by $q \equiv 1-1/2m^2L^2$. By a straightforward  near 
horizon analysis, one can see that
\bea && A(r) = \tilde{s}_0\CW(\phi_H)(r-r_H) +\cdots\,, \qquad  \qquad B(r) = 
\frac{1}{\CW(\phi_H)(r-r_H)} \cdots\,,  \\ \nn \\
       &&    \CW(\phi(r)) = \CW(\phi_H)  -\frac{1}{2}\CW(\phi_H)\Big[1- \frac{1}
{8m^2L^2}\CW^2(\phi_H)\Big]^{-1} (\phi-\phi_H)^2 +\cdots\,, \nn \\ \nn \\
       &&  \phi(r) = \phi_H + \tilde{g}_0(r-r_H)+ \cdots\,, \nn \eea
where $\tilde{s}_0$ and $\tilde{g}_0$  are   certain non-vanishing constants.  
One can show that the near horizon geometry becomes the so-called self-dual 
orbifold of $AdS_3$ space whose metric can be written as
\beq  ds^2_{NH} = \frac{\bar{L}^2}{4}\bigg[ -\rho^2dt^2  + \frac{1}{\rho^2}
d\rho^2 +4r^2_H\Big(d\theta - \frac{\rho}{2r_H}dt\Big)^2\bigg]\,, \qquad 
\bar{L} \equiv \frac{2L}{\CW(\phi_H)}\,.   \label{NHGeo}\eeq

In this case, it turns out that the $rt$ component of the potential is given by
\bea 
J^{rt} (\xi_R) &=&  \frac{1}{L^2}e^{-2B-A}F \bigg[ 1 + \frac{1}{2m^2L^2}
e^{-2B})''  -\frac{3}{4m^2L^2}\frac{F}{r} (e^{-2B})' \bigg]   \\  &&   
+\frac{3}{2m^2L^4}\, e^{-2B-A}\,F'\, (e^{-2B})'   ~~~ \nn ~~~ \\
&&
+ \frac{1}{8m^2L^4}e^{-4B-A} \bigg[   \frac{1}{r^2}F^3  -\frac{4}{r^2}F     
-\frac{12}{r}F\,F'    + 8 F'' \bigg] \,, \nn \\  \nn \\ 
  F &\equiv&  rA'(r) -1\,. \nn 
~~~ \eea
By using the eq.~(\ref{MFNMG}) and the eq.~(\ref{PsiNMG}), one can obtain a differential 
equation without the superpotential $\CW$. By combining the resultant equation with the
eq.~(\ref{MFNMGa}), one can show that  the  above quasi-local angular momentum is independent of the radial
 coordinate as
\beq  J_{\CB_r} (\xi_R) = \frac{L}{8G}\tilde{\Delta}\Big[1+ \frac{1}{2m^2L^2}\Big]  =   
 \frac{L}{8G}\tilde{\Delta}\Big[1 + \frac{1}{8m^2L^2}\CW^2(\phi_{\infty})\Big]      \,. \eeq
Note that the above expression of angular momentum at the asymptotic 
infinity for hairy deformed BTZ black holes is completely consistent with the previous results~\cite{Nam:2010ub}.  
It is interesting to compute the quasi-local angular momentum  directly on the 
near horizon geometry. The result is given by 
\beq  %J_{r\rightarrow \infty} &=&   \frac{L}{8G}\tilde{\Delta}\Big[1 + \frac{1}{2m^2L^2}\Big]\\ \nn \\ 
 J_{H} = \frac{L}{8G}~r^2_H\CW(\phi_H) \Big[1 + \frac{1}{8m^2L^2}\CW^2(\phi_H)\Big] \,.  \nn \eeq 
Now, by using the relation between $\tilde{\Delta}$ and $\CW(\phi_H)$ given in the eq.
(\ref{PsiNMG}), one can verify that
\beq J_{\CB_r} = J_{r\rightarrow \infty} = J_H\,. \eeq

Some comments are in order. Firstly, the $AdS$ radii  on the asymptotic infinity and 
on the horizon are different but give the same expression for the angular momentum.  
In the Einstein gravity limit this result is consistent with  the general argument given in the Appendix-B.  
Secondly, this result is also consistent with the one from the Brown-Henneaux 
method adopted  in \cite{Hotta:2008xt} for a specific example. Our result can be regarded 
as the generalization of this case to more generic extremal hairy black holes. 
Furthermore, we showed that the quasi-local angular momentum is invariant 
along whole RG trajectory not just at the two conformal points.  Finally, one may note that the invariance of angular momentum for the extremal 
hairy deformed BTZ black holes  is, more or less, connected with the reduced order 
EOM, which may have some implication for the entropy of 
black holes. 

Now we study the relationship between the entropy of the deformed extremal BTZ black holes  and holographic $c$-theorem.  
%For a Killing vector $\xi$,  its corresponding conserved charge may be  defined  through 
%the potential $J^{\mu\nu}(\xi)$ on the domain $B$ as
%%
%\beq Q_B(\xi) = \frac{1}{16\pi G}\int_{B} d \Sigma_{\mu\nu}\, J^{\mu\nu}(\xi) \,. \eeq
%%
%When the hyper surface $\Sigma$ has two boundaries $B_1$ and $B_2$, one can see  that 
% the quasi-local conserved charges at each boundary  are related as 
%%
%\beq     \int_{\CB_1} d \Sigma_{\mu\nu}\, J^{\mu\nu}(\xi) = \int_{\CB_2}  d \Sigma_{\mu\nu}\, 
%J^{\mu\nu}(\xi)  + \int_{\Sigma} d\Sigma_{\mu}\, J^{\mu}(\xi) \,. \eeq
%%
%This expression shows us that, whenever the current vanishes on the hypersurface 
%$\Sigma$,  the quais-local conserved charge is independent of its domain.  
%Interestingly, in our choice of the current $J^{\mu}$ satisfies this on-shell vanishing 
%property on the space-like hypersurface connecting the asymptotic infinity and the horizon.  
%Specifically for Einstein gravity coupled with scalar fields, our current is simply given by
%%
%\beq J^{\mu}(\xi) = 2R^{\mu\nu}\xi{\nu} \,. \eeq
%%
%For the rotational Killing vector $\xi_R$ the current becomes orthogonal to the hypersurface 
%$\Sigma$.  However, it seems not so trivial to check in the NMG case. 
The central  charge of the ultraviolet and the infrared dual CFT can be obtained through 
the dictionary of the AdS/CFT correspondence as
\bea c_{UV} &=&  \frac{3L}{2G}\Big[ 1 + \frac{1}{2m^2L^2} \Big]= \frac{3L}{G\CW(\phi_{\infty})}\Big[ 1 + \frac{1}{8m^2L^2} \CW^2(\phi_{\infty} ) \Big]  \,, \\      
 c_{IR} &=&   \frac{3\bar{L}}{2G}\Big[ 1 + \frac{1}{2m^2\bar{L}^2}  \Big]  = \frac{3L}{G\CW(\phi_H)}\Big[ 1 + \frac{1}{8m^2L^2} \CW^2(\phi_H) \Big]\,. \nn \eea
Though the central charge, $c_{UV}$,  of ultraviolet  dual CFT   has been derived in various ways~\cite{Nojiri:1999mh}, the central charge, $c_{IR}$  of the infrared dual CFT has not done explicitly. 
Following the discussion given at the end of the previous section, it is straightforward to obtain the expression for $c_{IR}$ for the extremal hairy 
BTZ black holes by using the near horizon geometry given in the eq. (\ref{NHGeo}). This would provide  a generalization of scalar-Einstein theory given in \cite{Hotta:2008xt}  to the NMG case.  
The holographic realization of the central charge function, which connects the above two central charges has been suggested in~\cite{Kwon:2012zh}.

By using the relations between total conserved charges and conformal weights of the ultraviolet dual CFT
\[
M_{\infty} = E_L + E_R\,, \qquad J_{\infty}  = L(E_L - E_R)\,, \]
one can see that the entropy of the ultraviolet CFT computed from the Cardy formula is given by
\beq S_{UV}  = 2\pi\sqrt{\frac{c_L}{6}(M_{\infty} L+J_{\infty} ) } + 2\pi \sqrt{\frac{c_R}{6}(M_{\infty} L - J_{\infty} )} = 2\pi\sqrt{\frac{c_{UV}}{3}J_{\infty} } \,, \eeq
where we have used the property of extremal black holes: $M_{\infty}L = J_{\infty}$.  
To apply the same relations for the infrared dual CFT on the near horizon self-dual orbifold of $AdS_3$, 
one needs the quasi-local expressions for mass and angular momentum. 
Using our result for the angular momentum 
invariance, $J_{\infty} = J_{H}$ and the extremality condition for conserved charges  $M_H\bar{L} = J_H$, 
it can be easily seen that  the entropy of the dual CFT is related to the BH entropy as  
\beq 
S_{UV}  \ge S_{IR} = S_{BH} = \frac{A_H}{4G}\Big[1 + \frac{1}{2m^2\bar{L}^2}\Big]\,,  \qquad A_{H} \equiv 2\pi \bar{L}r_H\,.  
\eeq
This result verifies our claim that the entropy of the infrared CFT is indeed the same as the 
usual black hole entropy.

%%%%%%%%%%%%%%%%%%%%%%%%%%%%%%%%%%%%%%%%%%%%%%%%%%
\section{Conclusion}
In this work we have studied the entropy and the invariance of the angular momentum for the rotating 
BTZ black holes in NMG by using the {\it off-shell} expressions for the Noether current and potential. We have also showed the invariance of the angular momentum for extremally rotating  scalar-hairy deformed BTZ black holes which can be interpreted as the RG flows for the dual field theory in the context of the AdS/CFT correspondence.

Firstly, we have computed the entropy for non-extremal rotating BTZ black hole by using the so-called 
stretched horizon approach.  In  this case, the entropy tensor $P^{abcd}$ is not divergence free. 
Consequently, the  generic expressions for {\it off-shell}  Noether current,   potential and conserved 
charge contain the terms proportional to $P^{abcd}$, $\nabla P$ and $\nabla \nabla P$.  
Simplification occurs when we evaluate these expressions in the vicinity of the Killing horizon. 
Near the Killing horizon, $\nabla P$ and $\nabla \nabla P$ terms are of the higher order in $\chi^{2}$ and 
hence does not contribute to the Virasoro algebra.  As a result, the final expression for 
the central term takes the same form with the one in  Einstein or Lanczos-Lovelock gravity 
\cite{Majhi:2011ws}. By Fourier-transforming this central extension term and 
using the ansatz for the scalar function $T$,  
we identified the central charge. Finally, by using this central charge and zero 
mode eigenvalue of the conserved Noether charges the black hole  entropy 
is obtained.  

We have also provided a brief derivation of the central charge for the rotating extremal 
BTZ case. In this case,  we also used the {\it off-shell} expressions for the Noether current and potential
as in the non-extremal case. However, we adopted the standard AdS/CFT  dictionary to obtain the central charge, for which we have taken the diffeomorphisms preserving
the fall-off boundary conditions at the asymptotic boundary of the near horizon extremal geometry.  Our result for the central charge is in agreement with the one given in \cite{Strominger:1997eq,Saida:1999ec} 
where the asymptotic boundary of whole extremal geometry is used. 
 
Secondly, we have established the invariance of the angular momentum along the radial direction for black holes in our models. 
To this purpose, 
we have used the same expressions for Noether current or potential and verified our claims explicitly. 
In the context of the AdS/CFT correspondence conserved charges of three-dimensional AdS black holes 
are related to the conformal weights of states dual to the black holes. 
Therefore, the angular momentum invariance 
along the radial direction in this set-up indicates a certain RG flow behavior of scaling dimensions of dual operators.  
This invariance also  plays a crucial role in the computation of the entropy. 
Our expression of the entropy for the extremally rotating BTZ black hole (see eq.(\ref{eq:entropyextremBTZ})) 
contains quasi-local angular momentum $J_{H}$, which corresponds to conformal weight in the infrared CFT. On the other hand, 
the Cardy formula for ultraviolate case  \cite{Strominger:1997eq,Saida:1999ec} uses the asymptotic value for the angular momentum, $J_{\infty}$, 
as the conformal weight. 
The fact that the angular momentum  is invariant along the radial direction allows us to 
match our result (\ref{eq:entropyextBTZ2}) with the corresponding one in the ultraviolate case. 
While this is somewhat anticipated by the stretched horizon approach to the black hole entropy, 
the quasi-local conserved charges need to be adopted to show this matching. 
Moreover, we have also shown the angular momentum invariance of  the deformed extremally rotating BTZ
black holes. In this case we verified that the corresponding entropy computed from the 
infrared CFT is less than the one from the ultraviolate case and it matches with the black hole entropy.  Since our boundary conditions are rather restricted, it would be interesting to study hairy deformed rotating AdS black holes with more generic boundary conditions and investigate the angular momentum invariance in those cases.

The angular momentum invariance discussed here is restricted to some specific black hole solutions. 
It would be interesting to extend this analysis to the more generic case. This would enable us to connect the 
generalized Komar potential used here and the ADT/ADM potentials for the generic higher derivative gravity. 
We would like to investigate these issues in the near future. 
 \section*{Acknowledgments}
This work was supported by the National Research Foundation (NRF) of Korea grant funded by the Korea government 
(MEST) through the Center for Quantum Spacetime (CQUeST) of Sogang University with grant number 2005-0049409.  
W. Kim was supported by Basic Science Research Program through the  National Research Foundation of Korea (NRF) funded by the Ministry of Education, Science and Technology (2010-0008359).
S.-H.Yi was supported by Basic Science 
Research Program through the NRF of Korea funded by the MEST (2012R1A1A2004410).
S. Kulkarni would like to thank B.R. Majhi for useful discussions. 
 S.-H. Yi would like to thank Seungjoon Hyun, Jahoon Jeong, Yongjoon Kwon, Soonkeon Nam and Jong-Dae Park for some discussions.

%In this case, the algebra among the {\it off-shell} Noether charges is obtained at the
%asymptotic infinity of the near horizon extramal BTZ black hole. The central charge identified
%from this algebra is actually the central charge for the infrared CFT.  
%
%derive the   
%We have considered rotating AdS black holes in NMG by using the modified {\it off-shell} 
%Noether current,  the potential of which is  coined   as the generalized Komar potential in this paper. 
% As an extension of this case, we considered the  extremal hairy AdS black holes in NMG and  have 
%showed that the angular momentum of these black holes is invariant. This result 
%verifies that the entropy of the near horizon dual CFT gives the BHW entropy of those black holes. 
%
%
%
%Because of the extremality of black hole solutions in our concern, mass and angular momentum of 
%black holes under consideration are not independent but proportional to each other. It is plausible to
% require  that this property should hold at the level of quasi-local conserved charges, not just at the
% level of total ones.

\newpage

%%%%%%%%%%%%%%%%%%%%%%%%%%%%%%%%%%%%%%%%%%%%%%%%%%
\section*{Appendix A: Some formulae in stretched horizon approach}
\renewcommand{\theequation}{A.\arabic{equation}}
\setcounter{equation}{0}
In this appendix we shall briefly state some results  in Carlip's stretched horizon 
approach\cite{Carlip:1999cy}. This will be useful in deriving the near horizon expressions
for Noether current, potential and charge. Detailed derivation of these results can be found in the Appendix-B of Ref. \cite{Majhi:2011ws}.

%Let us consider the $d+1$ dimensional Riemannian manifild. Given an approximate Killing vector $\chi^{a}$ in 
%the neighborhood of the boundary $\Sigma$ of the manifold, the local Killing horizon is defined by the condition $\chi^{2}=0$. In the stretched horizon approach, this is alternatively given by 
% \begin{equation}\label{eq:B1}
%\chi^{2} = \ep. 
%\end{equation}
%and $\ep$ is taken to be zero at the end. The vector orthogonal to the orbits of $\chi^{a}$ is
%given by
%\beq
%\rho_{a} = -\frac{1}{2\kappa}\nabla_{a}\chi^{2}~. \label{eq:B2} 
%\eeq
%Where $\kappa$ is the surface gravity associated with the Killing vector field $\chi^{a}$
%\beq
%\kappa^{2} \equiv \lim_{\chi^{2}\rightarrow 0 } \lb\frac{\nabla_{a}\chi^{2} \nabla^{a}\chi^{2}}{4\chi^{2}}\rb
%\eeq
%Consider a general diffeomorphism transformation 
%\beq
%\xi^{a} = T \chi^{a} + R \rho^{a} \label{eq:B3}
%\eeq
%Now we impose the condition that under such diffeomorphisms, 
%\beq
%\delta \chi^{2} = 0 \qquad ; \qquad  as  \qquad \chi^{2} \rightarrow 0 \label{eq:B4}
%\eeq
%This condition lead to a following relation between two arbitrary functions $T$ and $R$:
%\beq
%R = \frac{1}{\kappa}\frac{\chi^{2}}{\rho^{2}} D T \label{eq:B5}
%\eeq

 The approximate Killing vector $\chi^{a}$ and the vector  $\rho^{a}$ which is normal to  the Killing orbits satisfy
\bea
\frac{\rho^{2}}{\chi^{2}} &=& -1  \,,\label{eq:B6} \\
\nabla_{a}\chi_{b} &=& \frac{2\kappa}{\chi^{2}}\chi_{[a}\rho_{b]} \label{eq:B7} \,,\\
\nabla_{a}\rho_{b} &=&  \frac{\kappa}{\chi^{2}} (\chi_{a}\chi_{b}-\rho_{a}\rho_{b})\,. 
\label{eq:B9}
\eea
For the diffeomorphisms given by eq.(\ref{eq:gendiff}), we have
\bea
\nabla_{a}T &=& \frac{\chi_{a}}{\chi^{2}}DT \,,\label{eq:B8}\\
\nabla_{a}\xi_{b} &=& \frac{1}{\chi^{2}}\lb DT \rho_{a}\rho_{b} + 2\kappa T \chi_{[a}\rho_{b]} 
 - \frac{D^{2}T}{\kappa}\chi_{a}\rho_{b}\rb \label{eq:B10}\,,\\
\nabla_{a}\nabla_{b}\xi_{c} &=& \frac{1}{\chi^{4}}\lb \chi_{a}\chi_{b}\rho_{c}\ls2\kappa DT  - 
\frac{D^{3}T}{\kappa}\rs - D^{2}T \chi_{a}\chi_{b}\chi_{c}\rb~. \label{eq:B11}
\eea
These expressions are valid upto order $\chi^2$. 

\noindent Next, we give the expression for  $d-2$ dimensional surface element 
 $d\Sigma_{ab}$ :
\beq
d\Sigma_{ab} = \sqrt{h}d^{d-2}x  S_{ab}~, \label{eq:nullsurfaceelement}
\eeq
where $h$ is the determinent of the metric $h_{ab}$ on $d-2$ dimensional surface and 
\beq
S_{ab} = -\frac{2|\chi|}{\rho \chi^{2}}(\chi_{[a}\rho_{b]}) ~. \label{eq:S}
\eeq
Using the eq.(\ref{eq:S}) and the symmetries of $P^{abcd}$ one can easily verify the following identities:   

\beq
P^{abcd}\chi_{c}\rho_{d} =  -\frac{2|\chi|}{\rho \chi^{2}}P^{abcd}S_{cd}\,, \label{eq:Pidentity1}
\eeq

%\bea
%\nabla_{c}P^{abcd} S_{ab}\chi_{d} = -\frac{2|\chi|}{\rho \chi^{2}}\nabla_{c}P^{abcd}\chi_{a}\chi_{d}\rho_{b}
% \  \   ; \ \ \nabla_{c}P^{abcd} S_{ab} \rho_{d} = \frac{2|\chi|}{\rho \chi^{2}}\nabla_{c}P^{abcd}\rho_{a}\rho_{d}\chi_{b}
%\eea

\beq
\chi_{c}\chi_{e}\rho_{d}\rho_{b}P^{becd} = \frac{\rho^{2}\chi^{2}}{4}P^{becd}S_{be}S_{cd}~.\label{eq:Pidentity2}
\eeq
%%%%%%%%%%%%%%%%%%%%%%%%%%%%%%%%%%%%%%%%%%%%%%%%%%%
\renewcommand{\theequation}{B.\arabic{equation}}
  \setcounter{equation}{0}
\section*{Appendix B: Angular momentum invariance in Einstein gravity}
%%%%%%%%%%%%%%%%%%%%%%%%%%%%%%%%%%%%%%%%%%%%%%%%%%%
In this appendix we show  the angular momentum invariance in Einstein gravity minimally coupled with a scalar field. Note that our class of the hairy deformation of extremal BTZ black holes can be understood as the limit of the NMG case.  In this case  one can give more general argument for the angular momentum invariance.   

Through the so-called fake supersymmetry formalism, one may take the scalar potential in this case as~\cite{Hyun:2012bc}
\beq V(\phi)  = \frac{1}{2L^2}(\p_{\phi}\CW)^2 - \frac{1}{2L^2}\CW^2\,,\eeq
and it turns out that the extremally rotating AdS black holes can be described by the following first 
order EOM
\beq \phi' = -e^{B} \p_{\phi} \CW \,, \qquad A' = e^{B}\CW - \frac{1}{r}\,, \qquad A'+B' =
 \frac{r}{2}\phi'^2\,, \qquad (e^{C})' = \pm \Big(\frac{1}{r} e^{A}\Big)'\,, \eeq
where $'$ denotes the derivative with respect to the radial coordinate $r$. 
Note that the last equation can be integrate as 
 $e^{C(r)} =  C_{\pm} \pm e^{A(r)}/r$. Since the sign in this equation is related to the
 rotation direction, we will take the upper sign for definiteness. And the integration 
constant $C_{+}$ is taken as $C_{+} =1$ to match with the BTZ black holes asymptotically. 

By manipulating the first order EOM, one can show that the metric functions $A$ and $B$ are 
determined  in terms of a certain constant $\Delta$ and the superpotential $\CW$ as
\beq rA'(r) -1 = \frac{\Delta}{r}e^{B(r)}\,, \qquad e^{-B(r)} = \frac{r}{2}\Big[\CW - 
\frac{\Delta}{r^2}\Big]\,, \label{MetricFuc} \eeq
which correspond to $\Psi = \Delta =const.$  in the NMG case.

In this case of Einstein gravity,   it turns out that  all the relevant quantities, {\it e.g.}
the asymptotic and the  near horizon  expansion, the relation $\Delta =r^2_H\CW(\phi_H)$ and  the  Komar potential 
\beq J^{rt} (\xi_R) =  \frac{1}{L^2} e^{-2B(r)-A(r)}\Big[rA'(r) -1 \Big]\,, \eeq
can  be understood as the limit $m^2\rightarrow \infty$ in the NMG case. 

%where the prime denotes the differentiation with respect to $r$.  
By using the expression  for the metric function $A$ given in~(\ref{MetricFuc}) 
and by noting that $\sqrt{-\det g} = L^3 e^{A+B} r$, one can see that  the quasi-local angular momentum  is given by
\beq J_{\CB} =  \frac{L}{8G}\Delta=\frac{L}{8G}~ r^2_H\CW(\phi_H)= \frac{\bar{L}}{8G} \frac
{r^2_H\CW^2(\phi_H)}{2}\,, \eeq
where $\CB_r$ denotes the hypersurface of the constant radius $r$.  

One may apply our formalism for non-extremal hairy black holes. For rotating hairy 
AdS black holes given in~\cite{Correa:2012rc}, one can check that the Komar potential leads to
\beq  J^{rt}(\xi_R) = \frac{6}{L^2} \frac{\omega\, B^2(r+2B)^3}{(r+B)^4(1-\omega^2)}
\,.  \eeq
By using $\sqrt{-g} = L^3(r+B)^4/(r+2B)^3$ for the metric of those black holes, 
one can see that
\beq J_{\CB_r}= J_{\infty} = J_{H} = \frac{3L}{4G} \frac{\omega B^2}
{1-\omega^2}\,, \eeq
which is consistent with the result in~\cite{Correa:2012rc}  through the Hamiltonian formalism up  to the convention-dependent normalization. While these hairy AdS black holes satisfy more generic boundary conditions than Brown-Henneaux ones, the Komar integrand gives us the consistent result with the one given in~\cite{Correa:2012rc}.

In Einstein gravity, the more general argument for this angular momentum invariance along the radial direction can be given as follows.
When a hypersurface $\Sigma$ has two boundaries $\CB_1$ and $\CB_2$, one can see  that 
 the quasi-local conserved charges for a Killing vector $\xi$   at each boundary  are related as 
\beq     \int_{\CB_1} d \Sigma_{ab}\, J^{ab}(\xi) = \int_{\CB_2}  d \Sigma_{ab}\, 
J^{ab}(\xi)  + \int_{\Sigma} d\Sigma_{a}\, J^{a}(\xi) \,. \eeq
This expression shows us that, whenever the current, $J^a$ vanishes on the hypersurface 
$\Sigma$,  the quais-local conserved charge is independent of its domain.  
Interestingly for Einstein gravity coupled with scalar fields, our current for a Killing vector $\xi$ is simply given by
\beq J^{a}(\xi) = 2P^{abcd}\nabla_{b}\nabla_{c}\xi_d = 2R^{\mu\nu}\xi{\nu} \,. \eeq
The angular momentum invariance follows from the fact that, for the rotational Killing vector $\xi_R$,  the hypersurface $\Sigma$  connecting two boundary $\CB_{r=r_H}$ and $\CB_{r=\infty}$  can be chosen such that it is orthogonal to the current, $J^a(\xi_R)$. 
%It is  not so clear how one may apply this argument in the NMG case. 
%%%%%%%%%%%%%%%%%%%%%%%%%%%%%%%%%%%%%%%%%%%%%

%%%%%%%%%%%%%%%%%%%%%%%%%%%%%%%%%%%%%%%%%%%%%

 %%%%%%%%%%%%%%%%%%%%%%%%%%%%%%%%%%%%%%%%%%%%%%%%%%%%%%%%%%%%%%%%%%%%%%%%%%%%%%%%%%%%%%%%%%%%%%%%%%%%%%%%%%%%%%%%%%%%%%%%%%%%%%
\end{document}